\documentclass[aps,showpacs]{revtex4}

\usepackage{makeidx}
\usepackage{graphicx}

\begin{document}

\title{Integrability and chaos: the classical uncertainty}
\author{Jaume Masoliver$^1$, Ana Ros$^2$}
\affiliation{$^1$ Departament de F\'{\i}sica Fonamental. Universitat de Barcelona.\\Diagonal 647, 08028 Barcelona, Spain.}
\affiliation{$^2$ Department Algebra und Zahlentheorie. Universit\"at Hamburg. \\ Bundestrasse 55, 20146 Hamburg, Germany.}
\email{jaume.masoliver@ub.edu, anaroscamacho@gmail.com}

\begin{abstract}
In recent years there has been a considerable increase in the publishing of textbooks and monographs covering what was formerly known as random or irregular deterministic motion, now named by the more fashionable term of deterministic chaos. There is still substantial interest in a matter that is included in many graduate and even undergraduate courses on classical mechanics. Based on the Hamiltonian formalism, the main objective of this article is to provide, from the physicist's point of view, an overall and intuitive review of this broad subject (with some emphasis on the KAM theorem and the stability of planetary motions) which may be useful to both students and instructors.    
\end{abstract}
\maketitle

\section{Introduction}

At the beginning of the 20th century ``mechanics had not yet lost its place as the central subject within physics, although it was on the verge of losing it'' \cite{klein}. Fifty years later classical mechanics was considered, by the majority of physicists, a closed field from which one could not expect new challenges, specially in fundamental issues. But from then on, fresh  mathematical results, together with additional numerical data obtained by high-speed computers, revived a field of knowledge related to non-linear dynamics and irregular behaviour of deterministic systems. A field that had begun at the end of the 19th century mainly by the works of Henri Poincar{\'e} in his attempt to develop a convergent perturbation theory of planetary orbits.

The solar system, despite its complexity, shows an extremely regular behaviour which can be predicted very accurately for short periods of time. This is due, in part, to the weakness of the gravitational force, but also because Kepler's two-body problem is completely integrable, despite the fact that any gravitational system of three or more bodies is not (let us imagine for a moment the difficulties of guessing the law of gravitation had the Earth been orbiting around a double star). The Newtonian deduction of Kepler's laws is based precisely on the properties of the integrable system of two bodies. Nevertheless, the dynamical interactions of the many bodies that form the solar system necessarily lead us to consider deviations to the predictions based on Kepler's laws. This makes us wonder: why does the solar system behave in such a regular way? Or, as in the inquiry of J. K. Moser: is the solar system stable? \cite{moser(a)} That is, will it follow in the future the order that we presently observe? Even today these questions do not have a complete and satisfactory answer.

For more than three hundred years, since the time of Newton, the evolution and stability of mechanical systems, and specially that of the solar system, have been central to physics and mathematics. During the 18th century, Euler, Lagrange and Laplace made substantial developments by predicting changes in the planetary orbits due to small perturbations, establishing a framework for the study of global stability. All of this culminated in the 19th century works of Hamilton and Jacobi who reformulated the Lagrangian formalism of classical mechanics in terms of the phase space, a step which proved to be most fruitful for the developments of both statistical and quantum mechanics. 

These successes consolidated the idea of classical determinism. However, at the turn of the 19th century, the works of Poincar\'e not only closed the door to an age but generated the first serious fracture in the philosophical conception of determinism. These works showed the impossibility of proving the convergence of perturbation series and, therefore, central questions, as the stability of the solar system, remained unanswered. Poincar\' e was thus the first to study what much later has been called (apparently by J. Yorke in 1975) {\it chaos}. 

A plain, albeit not simplistic, definition of chaos is ``chaos is the complicated temporal evolution of simple systems'' \cite{tel}. Thus, and contrary to common thinking, chaos is not spatial and static disorder but a characteristic of certain motions. It is essentially a dynamical concept, in which two initially close trajectories may be divergent after a finite time. Chaos may appear in a variety of fields: physics, chemical reactions, the spread of illnesses, Internet, economics and, of course, in planetary motions \cite{tel}. For contrary to intuition, chaotic behaviour is not an exclusive property of dissipative or random systems but of conservative and deterministic systems as well.   

Focusing on conservative systems we will show that small denominators giving place to resonances are associated to the appearance of chaos, and that due to this irregular behaviour, it seems impossible to predict the long-term evolution of nonintegrable systems (the vast majority of mechanical systems) using perturbation techniques. Since the end of the 19th century there was no remarkable advance and this fundamental question was practically forgotten until 1954, when A. N. Kolmogorov gave a sketch of the proof that most trajectories of nonintegrable conservative systems are quasiperiodic and that one can obtain them through a convergent perturbation development. In the 1960's, V. I. Arnold and J. K. Moser gave a more formal and rigorous proof of this result, known as the Kolmogorov-Arnold-Moser (KAM) Theorem. 

In the coming sections we will explain and develop these ideas, stressing physical aspects, rather than mathematical concepts and technicalities, of which there are plenty in a field that was the starting point of modern mathematical developments such as topology. In this way the paper is intended as a comprehensive and intuitive introduction to deterministic chaos which, we hope, will be useful for either students and instructors.

\section{Hamiltonian mechanics}

The quintessential mechanical system, to which most physical systems are reduced, consists in a collection of mass points interacting between them. Experience shows that the state of the system is completely determined by the set of positions and velocities of all its particles. The reference system chosen to describe positions and velocities needs not be Cartesian. Thus, in the Lagrangian formulation of mechanics, the representation of the state of the system is obtained through generalized coordinates $q_1,\dots,q_n$ and velocities $\dot{q}_1,\dots,\dot{q}_n$. The minimum number of generalized coordinates needed to describe the state of the system is called the number of degrees of freedom. In what follows we will use bold letters to design vectors. Accordingly, the set of all positions is represented by the vector $\mathbf{q}=\left( q_1,\dots,q_n \right)$ and the set of all velocities by $\mathbf{\dot{q}}=(\dot{q}_1,\dots,\dot{q}_n)$. 

The laws of mechanics are those that determine the motion of the system by providing the time evolution of the positions, $\mathbf{q}(t)$, and the velocities, $\mathbf{\dot{q}}(t)$, of all its components. Therefore, the main problem is the determination of how the state of the system evolves with time. There have been several approaches to achieve this goal. The first one was based on Newton's laws which directly emanate from experience. This approach was subsequently refined and generalized by the Lagrangian and Hamiltonian formulations, being the latter the most appropriate for the present development.  

In the Hamiltonian formulation, the description of the state the system is also given by generalized coordinates $\mathbf{q}=\left( q_1,\dots,q_n \right)$ but instead of velocities by  generalized momenta $\mathbf{p}=\left( p_1,\dots,p_n \right)$. The main problem is then finding the time evolution of these quantities, in other words, the knowledge of the functions $\mathbf{q}(t)$ and $\mathbf{p}(t)$ which give the temporal evolution of coordinates and momenta. This is achieved by solving the following set of $2n$ first-order differential equations known as {\it Hamilton equations} \cite{born,goldstein,landau}:
\begin{equation}
\mathbf{\dot{q}}=\frac{\partial H}{\partial \mathbf{p}},\qquad
\mathbf{\dot{p}}=
-\frac{\partial H}{\partial \mathbf{q}},
\label{1-4}
\end{equation}
where the function $H \left( \mathbf{q},\mathbf{p},t \right)$ is the Hamiltonian which, for inertial reference frames, coincides with the total energy of the system \cite{born}.
In (\ref{1-4}) we have used the notation: 
$$
\frac{\partial H}{\partial \mathbf{q}}=\left( \frac{\partial H}{\partial q_1},\dots, \frac{\partial H}{\partial q_n} \right)
$$
to denote the gradient of $H$ with respect to $\mathbf{q}$ and analogously for $\partial H/\partial \mathbf{p}$.

For those readers not familiar with the Hamiltonian formulation, it may be useful to recall that for the three dimensional motion of a particle of mass $m$ inside a conservative field of potential energy $V(\mathbf{r})$, the energy (i.e., the Hamiltonian) is 
$$
H=\frac{p^2}{2m}+V(\mathbf{r}),
$$
where $\mathbf{q}=\mathbf{r}=(x,y,z)$ and $p^2=\mathbf{p}\cdot\mathbf{p}=p_x^2+p_y^2+p_z^2$. In this case Hamilton equations (\ref{1-4}) read
$$
\mathbf{\dot{r}}=\frac{\mathbf{p}}{m},\qquad \mathbf{\dot{p}}=-\frac{\partial V}{\partial\mathbf{r}},
$$
which coincide with Newton's second law.  

In the Hamiltonian formulation, the state of a system with $n$ degrees of freedom is described by a single point $(\mathbf{q},\mathbf{p})=(q_1,\dots,q_n,p_1,\dots,p_n)$ of a space of dimension $2n$ called \textit{phase space}. Hence, starting at $t=0$ from an initial point $\left( \mathbf{q}_0, \mathbf{p}_0 \right)$, the state of the system in a later time $t$ will be given by the point $\left( \mathbf{q} \left( t \right), \mathbf{p} \left( t \right) \right)$, where $\mathbf{q} \left( t \right)$ and $\mathbf{p} \left( t \right)$ are the equations of motion, that is, the solutions of Hamilton equations (\ref{1-4}) with the initial conditions $\mathbf{q} \left( 0 \right)=\mathbf{q}_0$ and $\mathbf{p} \left( 0 \right)=\mathbf{p}_0$. The trajectory described by the point $\left(\mathbf{q}(t),\mathbf{p}(t)\right)$ as time progresses is called the {\it phase-space trajectory} of the system.

One of the advantages of the Hamiltonian formulation is that we can evaluate the time evolution of any property of the system without having to know the form of the equations of motion. That is to say, without solving Hamilton equations which is usually quite involved. Let us assume that a given property of the system (for instance, the energy or the angular momentum) can be represented by a function, $f(\mathbf{q}, \mathbf{p},t)$, of the state of the system and the time. How does $f$ change when the state $\left( \mathbf{q}(t),\mathbf{p}(t) \right)$ evolves with time? The answer is simple and reads \cite{goldstein,landau}
\begin{equation}
\frac{d f}{d t}=\frac{\partial f}{\partial t}+\left[ f,H \right],
\label{1-5}
\end{equation}
where $\left[ f,H \right]$ is the \textit{Poisson bracket} defined by
\begin{equation}
\left[ f,H \right]=\sum_{i=1}^n\left(\frac{\partial f}{\partial q_i}
\frac{\partial H}{\partial p_i}-\frac{\partial f}{\partial p_i}
\frac{\partial H}{\partial q_i}\right).
\label{1-6}
\end{equation}

Observe that if $f$ does not depend explicitly of time, $\partial f/\partial t=0$ and $f$ will be constant (i.e., $d f/d t=0$) if its Poisson bracket with the Hamiltonian vanishes: 
$[f,H]=0.$ In such a case $f(\mathbf{q},\mathbf{p})$ is a {\it constant of motion}. Incidentally, for conservative systems the Hamiltonian is time independent, hence $\partial H/\partial t=0$ and, since $[H,H]=0$, we see that $H$ is a constant of motion, the energy.

The Poisson bracket may be extended to any pair of dynamical functions $f(\mathbf{q},\mathbf{p},t)$ and $g(\mathbf{q},\mathbf{p},t)$. A direct generalization of (\ref{1-6}) gives
\begin{equation}
[f,g]=\sum_{i=1}^n\left(\frac{\partial f}{\partial q_i}
\frac{\partial g}{\partial p_i}-\frac{\partial f}{\partial p_i}
\frac{\partial g}{\partial q_i}\right).
\label{poisson}
\end{equation}
From this definition one easily obtains a series of properties satisfied by the Poisson bracket. Properties that endow the set of dynamical functions with an algebraic structure called {\it Lie algebra}. We will not proceed here with this mathematical development and address the interested reader to specialized texts (see, for instance, \cite{fuchs}). 

For the special case that $f$ and $g$ are any of the $q$'s or $p$'s, definition (\ref{poisson}) gives 
\begin{equation}
[q_i,q_j]=[p_i,p_j]=0, \qquad [q_i,p_j]=\delta_{ij}.
\label{canonical}
\end{equation}
This is the condition for any set of variables $(\mathbf{p},\mathbf{q})$ to be a canonical set, i.e., that they satisfy Hamilton equations with some Hamiltonian $H$ \cite{goldstein}.

\subsection{Canonical transformations. The Hamilton-Jacobi equation}

Perhaps the greatest advantage of the Hamiltonian formulation is that it allows for transformations of variables preserving the form of Hamilton equations. There are several ways, all of them equivalent, to define these kind of transformations. We choose the following:

We say that the change of variables, $\left( \mathbf{q},\mathbf{p} \right) \rightarrow \left( {\mathbf{q'}}, {\mathbf{p'}} \right)$, given by the equations 
\begin{equation}
\mathbf{q}=\mathbf{q} \left({\mathbf{q'}}, {\mathbf{p'}},t \right), \qquad
\mathbf{p}=\mathbf{p} \left({\mathbf{q'}}, {\mathbf{p'}},t \right),
\label{canonical_1}
\end{equation}
is a \textit{canonical transformation} if the infinitesimal quantities ${\mathbf{p'}}\cdot d {\mathbf{q'}}=\sum {p}'_id {q}'_i$ and $\mathbf{p}\cdot d\mathbf{q}=\sum p_idq_i$ differ only in the total differential of a scalar function $S$:
\begin{equation}
{\mathbf{p'}}\cdot d{\mathbf{q'}}-\mathbf{p}\cdot d\mathbf{q}=dS.
\label{canonical_2}
\end{equation}
The function $S$ is called the {\it generating function} of the canonical transformation. 

Observe that, because of (\ref{canonical_1}), out of the four sets of variables: $\mathbf{q},\mathbf{p},{\mathbf{q'}}$ and ${\mathbf{p'}}$, involved in a canonical transformation, only two of them are actually independent. The generating function $S$ will thus depend upon the two sets of variables chosen to be independent, and also on $t$ in the case of a time-dependent canonical transformation. One convenient choice, among others, consists in assuming that the generating function depends on the original coordinates and the transformed momenta, i.e,  $S=S(\mathbf{q},\mathbf{p'},t).$ 
Combining this assumption with condition (\ref{canonical_2}) one can easily show that the equations of the canonical transformation are given by  \cite{goldstein}
\begin{equation}
\mathbf{p}=\frac{\partial S}{\partial\mathbf{q}}, \qquad \mathbf{q'}=\frac{\partial S}{\partial\mathbf{p'}}.
\label{canonical_3}
\end{equation}

It can be shown \cite{goldstein,landau} that after a canonical transformation the Hamilton equations preserve the form:
\begin{equation}
\mathbf{\dot{q}}'=\frac{\partial H'}{\partial \mathbf{p}'},\qquad
\mathbf{\dot{p}}'=-\frac{\partial H'}{\partial \mathbf{q}'},
\label{hamilton_trans}
\end{equation}
where
\begin{equation}
{H}'\left( {\mathbf{q'}},{\mathbf{p'}},t \right)=H \left( \mathbf{q},\mathbf{p},t \right)+ \frac{\partial S}{\partial t},
\label{1-6a}
\end{equation}
is the transformed Hamiltonian. Note that for time-independent canonical transformations $\partial S/\partial t=0$, and $H'=H$, so that the transformed Hamiltonian coincides with the original one (the latter being written in terms of the transformed variables). 

Let us observe that {\it any} arbitrary function $S$ generates {\it some} canonical transformation. This is quite remarkable, because it implies the freedom of writing Hamilton equations --that is, the laws of motion-- in any convenient way only by the introduction of a given function $S$. 

Following this line of reasoning we might look for a canonical transformation such that the equations of motion would be the simplest ones. In other words, what would it be the generating function of the canonical transformation such that the transformed $q'$s and $p'$s are constants? Note that, if we could find this transformation, then we would have solved the problem of motion, for the equations of the transformation (\ref{canonical_1}) (or, equivalently, (\ref{canonical_3})) give the equations of motion $\mathbf{q}(t)=\mathbf{q}(\mbox{\boldmath$\alpha$},\mbox{\boldmath$\beta$},t)$ and $\mathbf{p}(t)=\mathbf{p}(\mbox{\boldmath$\alpha$},\mbox{\boldmath$\beta$},t)$, where the new coordinates and momenta $\mathbf{q`}=\mbox{\boldmath$\alpha$}$ and $\mathbf{p`}=\mbox{\boldmath$\beta$}$ are constants determined by the initial conditions. 

If the transformed variables are constant, their time derivatives are zero, and from (\ref{hamilton_trans}) we see that 
$\partial H'/\partial \mathbf{p}'=\partial H'/\partial \mathbf{q}'=0$. Accordingly, the transformed Hamiltonian $H'$ must be independent of the coordinates and momenta and, at most, a function of time, which, for simplicity, we choose to be zero. Therefore, if in (\ref{1-6a}) we set $H'=0$, and remembering that 
$\mathbf{p}=\partial S/\partial\mathbf{q}$ (cf (\ref{canonical_3})), we see that the generating function of the canonical transformation 
$(\mathbf{q},\mathbf{p}) \rightarrow (\mbox{\boldmath$\alpha$},\mbox{\boldmath$\beta$})$ obeys the equation
\begin{equation}
\frac{\partial S}{\partial t}+H\left(\mathbf{q},\frac{\partial S}{\partial\mathbf{q}},t\right)=0
\label{hamilton-jacobi}
\end{equation}
called the {\it Hamilton-Jacobi equation}. This is a first-order nonlinear partial differential equation and it provides an alternative procedure for finding the equations of motion  \cite{born,goldstein,landau}.

\section{Integrable Hamiltonians}

In classical mechanics one of the most fundamental questions is the {\it integrability} of a given system. Here integrability does not mean whether or not the Hamilton equations can be solved in some particular way. In fact, as a set of $2n$ differential equations, they can always be solved (at least numerically) under very general conditions given by the Cauchy theorem \cite{whittaker}.\footnote{The same considerations apply to the alternative procedure provided by the Hamilton-Jacobi equation.}

The concept of integrability refers to {\it the existence of constants of motion} (also called {\it integrals} and hence the name) which are responsible for the regular evolution of the phase-space trajectories of the system in well-defined regions of phase space. Due to its importance, let us recall again that a constant of motion is any function of positions and momenta which does not change with time whereas the $q$'s and $p$'s do.

There are several equivalent definitions of integrability \cite{goldstein,rasband,tabor,whittaker}, all of them underlying, in one way or another, the fact that {\it Hamilton equations (or the Hamilton-Jacobi equation) are integrated in terms of as many constants of motion as degrees of freedom}. One of these definitions, which turns out to be convenient for our purposes, is the following: 

\begin{quotation}

\noindent
{\it The Hamiltonian of a system with $n$ degrees of freedom is integrable if there exist $n$ independent constants, or integrals of motion, 
$I_i \left( \mathbf{q},\mathbf{p} \right)=\alpha_i$ $(\alpha_i={\rm constant},\ i=1,\dots,n)$, which are in ``involution'', meaning that the quantities $I_i$ satisfy
\begin{equation}
\left[ I_i,I_j \right]=0
\label{2-0}
\end{equation}
$(i,j=1,2,\dots,n)$ where $\left[ I_i,I_j \right]$ is the Poisson bracket defined as in} (\ref{poisson}).

\end{quotation}

\noindent
Let us remark that the mere existence of $n$ independent constants of motions is not sufficient for integrability; to ensure that any system is integrable it is also necessary that the constants of motion satisfy the property of involution (\ref{2-0}). Property that guarantees the existence of a canonical transformation for which the transformed momenta are precisely the integrals of motion $p_i'=I_i$ and, therefore, constants. For constant momenta the integration of Hamilton equations is immediate. This is essentially the content of a theorem attributed to Liouville \cite{whittaker} and what ultimately justifies the above definition of integrability (see \ref{appenA0} for a more complete discussion).

Unfortunately, the existence of $n$ integrals of motion in involution is the exception more than the rule. Poincar{\'e} proved, more than a century ago, that most mechanical systems only have the energy as a constant of motion. Therefore, {\it most systems with more than one degree of freedom are not integrable}. 

A paradigmatic example is provided by a closed $N$-body system, i.e., a set of $N$ point particles interacting with each other without external forces acting on them.\footnote{This is the case, within certain approximation, of the Solar System, the Milky Way or even the system Moon-Earth-Sun among many others.} The $N$ body system has $3N$ degrees of freedom ($3$ for each particle) and only $10$ integrals of motion. These are: 

\begin{itemize}

\item The three components of the total momentum
$$
\mathbf{P}=\sum_{k=1}^N\mathbf{p_k}=\ {\rm constant}.
$$

\item The three components of the motion of the center of mass ($M$ is the total mass of the system)
$$
\mathbf{R}_{CM}(t)-\frac 1M\mathbf{P}t=\ {\rm constant}.
$$

\item The three components of the total angular momentum
$$
\mathbf{L}=\sum_{k=1}^N\mathbf{l_k}=\mathbf{R}_{CM}\times\mathbf{P}+\mathbf{l}=\ {\rm constant},
$$
where $\mathbf{l}$ is the relative angular momentum. 

\item The energy
$$
E=\frac{1}{2M}\mathbf{P}^2+T_{rel}+V=\ {\rm constant}.
$$
where $T_{rel}$ is the relative kinetic energy and $V$ is the potential energy. 

\end{itemize}

\noindent
From these ten integrals {\it only six of them are independent and in involution}: the three components of the total momentum $\mathbf{P}$, the square of the relative angular momentum $\vert \mathbf{l} \vert^2$, the third component of the relative angular momentum $l_3$, and the total relative energy $T_{rel}+V$ \cite{scheck}.

When $N=2$, {\it the two-body problem}, there are $6$ degrees of freedom and the problem is integrable. However, when $N=3$, {\it the three-body problem}, there are $9$ degrees of freedom but only $6$ independent constants of motion in involution; whence, the three-body problem (for example, the system Sun-Earth-Moon) is not integrable.

\subsection{Invariant tori}

Let us first focus on integrable systems and postpone nonintegrable systems for later sections. A first and most important consequence of the existence of $n$ involution constants $I_i$ is that the trajectories of the system cannot pass through all points of the phase space. This is due to the fact that all trajectories are confined to move in a $n$ dimensional manifold $M$ determined by the constants of motion and embedded in the phase space of dimension $2n$. Indeed, the manifold $M$ is defined by the set of $n$ algebraic equations:
\begin{equation}
I_1(\mathbf{q},{\bf p})=\alpha_1 \quad \dots \quad I_n(\mathbf{q},{\bf p})=\alpha_n, 
\label{integrals}
\end{equation}
where $\alpha_1, \cdots \alpha_n$ are arbitrary constants. Consequently, {\it the phase-space trajectories never leave} $M$ because otherwise the conservation laws would be violated. Note that the manifold $M$ depends on the particular value of the constants of motion and, since the latter depend on the choice of the initial state of the system, we see that $M$ is fixed by the initial conditions. 

In what follows we will assume that the motion of the system is bounded (recall, for instance, the planetary motion). As a consequence of this, the manifold $M$ must be bounded as well. In fact, it can be shown, that $M$ is a compact (closed and bounded) manifold \cite{rasband}. We will now see that in this case $M$ {\it has the topology of a $n$ dimensional torus}. 

In effect, let us define the following $n$ velocity fields:
\begin{equation}
\mathbf{v}_i\equiv
\left(\frac{\partial I_i}{\partial\mathbf{p}},
-\frac{\partial I_i}{\partial\mathbf{q}}\right),
\label{2-1}
\end{equation}
($i=1,2,\dots,n$). Note that if, for example, $I_1=H$ then (cf (\ref{1-4})) $\mathbf{v}_1=(\mathbf{\dot{q}},\mathbf{\dot{p}})$. Accordingly, in (\ref{2-1}) is included the Hamiltonian flux of the phase-space trajectories which, due to the existence of $n$ independent integrals in involution, must be completely contained in the manifold $M$. 

In order to visualize that $M$ has the topology of an $n$ dimensional torus, let us first show that the fields $\mathbf{v}_i$ ($i=1,2,\dots,n$) are independent and tangent to $M$. The property of independence immediately follows from the independence of the integrals $I_i$. We next observe that the $2n$ dimensional gradients:
$$
\mbox{\boldmath$\nabla$}I_j=\left(\frac{\partial I_j}{\partial\mathbf{q}},\frac{\partial I_j}{\partial\mathbf{p}}\right),
$$
($j=1,\dots,n$), are orthogonal to every surface $I_j(\mathbf{p},{\bf q})=\alpha_j$. 

On the other hand, the scalar product of the fields $\mathbf{v}_i$ with these gradients is
$$
\mathbf{v}_i\cdot\mbox{\boldmath$\nabla$}I_j=\frac{\partial I_i}{\partial\mathbf{p}}\cdot\frac{\partial I_j}{\partial\mathbf{q}}-
\frac{\partial I_i}{\partial\mathbf{q}}\cdot\frac{\partial I_j}{\partial\mathbf{p}}=
\sum_{k=1}^{n}\Biggl(\frac{\partial I_i}{\partial q_k}\frac{\partial I_j}{\partial p_k}-
\frac{\partial I_i}{\partial p_k}\frac{\partial I_j}{\partial q_k}\Biggr)
$$
and, by virtue of the involution property, we get
$$
\mathbf{v}_i\cdot\mbox{\boldmath$\nabla$}I_j=\left[I_i,I_j\right]=0 
$$
$(i,j=1,2,\dots,n)$. Hence, the fields $\mathbf{v}_i$ are orthogonal to $\mbox{\boldmath$\nabla$}I_j$ and, since the gradients $\mbox{\boldmath$\nabla$}I_j$ are orthogonal to $M$, we conclude that {\it the fields $\mathbf{v}_i$ are tangent to the manifold $M$}. 

\begin{figure}
\begin{center}
\includegraphics[scale=0.70]{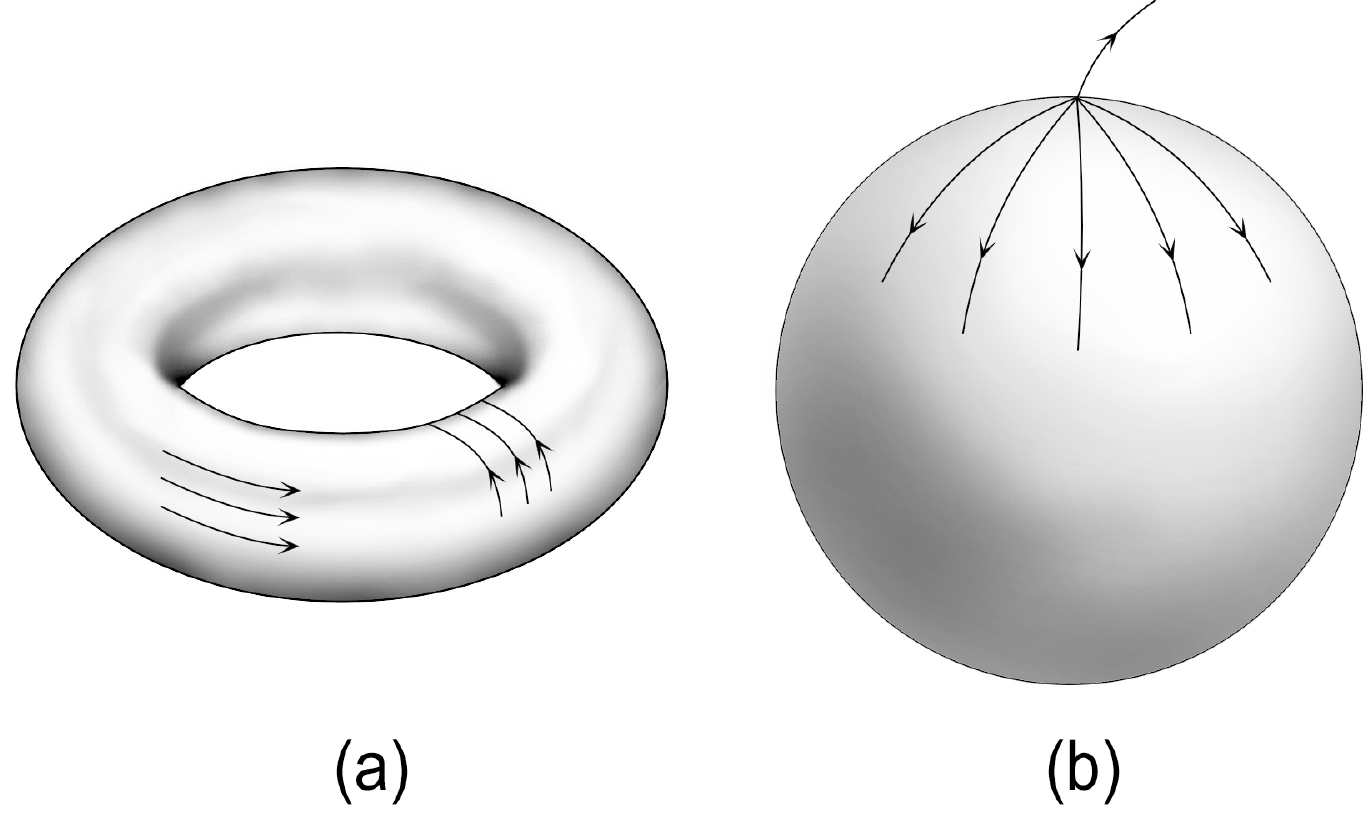} 
\caption{The ``hairy ball theorem''. (a) Vector fields smoothly ``combed'' on a $2$ dimensional torus. (b) Singular point of a vector field on a $2$ dimensional sphere.}
\label{Fig1}
\end{center}
\end{figure}

There is a theorem in topology --the Poincar{\'e}-Hopf theorem, jokingly known as the ``hairy ball theorem''-- which states that if in a compact manifold $M$ of dimension $n$, one can define $n$ independent vector fields tangent to $M$, then $M$ has necessarily the topology of an $n$ dimensional torus \cite{arnold,rasband,tabor}. 

Without getting into technical details, we can intuitively imagine this result as follows. Among the several possibilities for a manifold $M$ to be compact, and hence bounded, the simplest ones are: either $M$ is a $n$ dimensional sphere (or a surface with the same topology, like an apple) or it has the topology of an $n$ dimensional torus. However, it is impossible to define a vector field on a sphere in all its points while it is perfectly possible to define a vector field on the surface of a torus. This is easily visualized if we compare the act of combing down hair on a torus and on a sphere --for the latter case, a hair always sticks up at the pole and, hence, it is not tangent to the sphere--  see Fig. \ref{Fig1}.

\section{Action-angle variables}

We have just seen that, for the bounded motion of integrable systems with $n$ degrees of freedom, the phase-space trajectories lie on $n$ dimensional tori embedded in the $2n$ dimensional phase space. We will now see that the existence of these invariant tori\footnote{The tori defined in (\ref{integrals}) are often called ``invariant tori'' because they are constructed by giving different values to the integral invariants $I_i$.}  provides a way of defining a new set of canonical variables, the action-angle variables, which are crucial for the further development of the subject. 

\begin{figure}
\begin{center}
\includegraphics[scale=0.70]{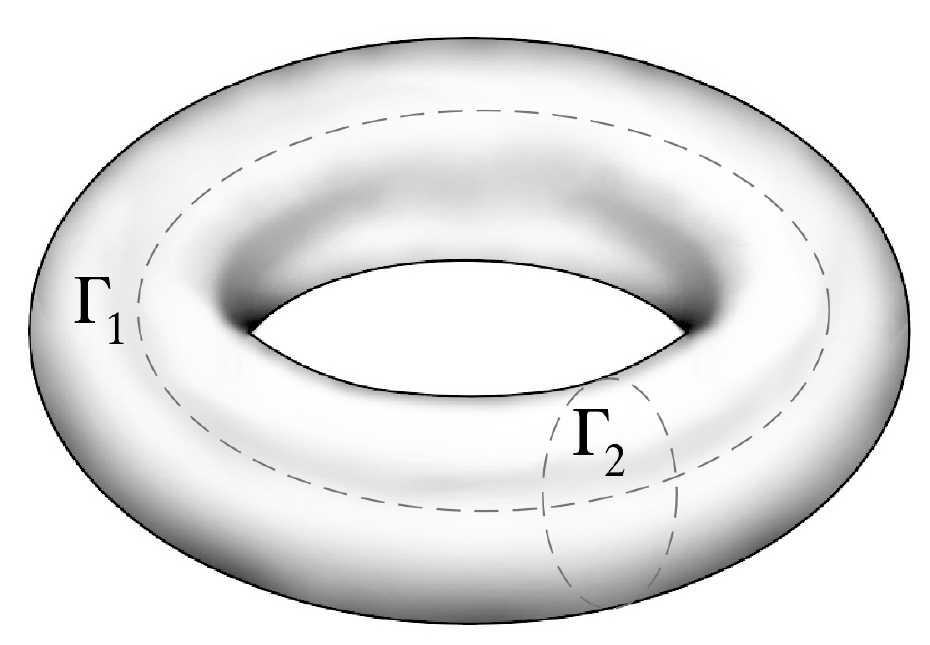} 
\caption{The two topological independent curves $\Gamma_1$ and $\Gamma_2$ on a $2$ dimensional torus.}
\label{Fig2}
\end{center}
\end{figure}

Indeed, any $n$ dimensional torus is a periodic object which can be considered as the product of $n$ independent periodicities. In other words, we can define $n$ topologically independent closed curves, $\Gamma_1,\dots,\Gamma_n$, on a given torus, where none of the $\Gamma_i$ can be deformed continuously into each other or shrunk to zero \cite{rasband,tabor}. In Fig. \ref{Fig2} we represent a $2$ dimensional torus for which $\Gamma_1$ turns around through the longest path while $\Gamma_2$ does it through the shortest path. Note that, neither $\Gamma_1$ nor $\Gamma_2$ can be converted into each other by continuous transformations.  

Having this set of closed curves at our disposal, we define the \textit{action variables} by means of the line integrals:
\begin{equation}
J_i=\frac{1}{2\pi}\oint_{\Gamma_i}\sum_{k=1}^np_kdq_k,
\label{2-2}
\end{equation}
where 
\begin{equation}
p_k=p_k \left(\mathbf{q},\mbox{\boldmath$\alpha$} \right)
\label{momenta0}
\end{equation}
are the momenta written in terms of the coordinates $\mathbf{q}=\left( q_1,\dots,q_n \right)$ and the constants of motion $\mbox{\boldmath$\alpha$}=\left( \alpha_1,\dots,\alpha_n \right)$. Notice that (\ref{momenta0}) is obtained by solving for the $p$'s the set of algebraic equations (\ref{integrals}) provided by the existence of $n$ independent integrals. 

Since in (\ref{2-2}) we integrate with respect to the generalized coordinates, the action variables only depend on the integral invariants: 
\begin{equation}
J_i=J_i \left( \alpha_1,\dots,\alpha_n \right),
\label{J_alpha}
\end{equation}
$(i=1,\dots,n)$, and are themselves constants of motion.

We now define a canonical transformation from the original variables $\left( \mathbf{q},\mathbf{p} \right)$ to a new set of variables 
$\left( \mbox{\boldmath$\theta$},\mathbf{J} \right)$ where the actions are the transformed momenta. A generating function for this transformation is given by the indefinite integral
\begin{equation}
S \left( \mathbf{q},\mathbf{J} \right)=\int\sum_{k=1}^np_k(\mathbf{q},\mathbf{J})dq_k,
\label{S}
\end{equation}
where $p_k(\mathbf{q},\mathbf{J})$ is given by (\ref{momenta0}) in which the $\alpha$'s have been written in terms of the $J$'s after solving (\ref{J_alpha}). 

\begin{figure}
\begin{center}
\includegraphics[scale=0.70]{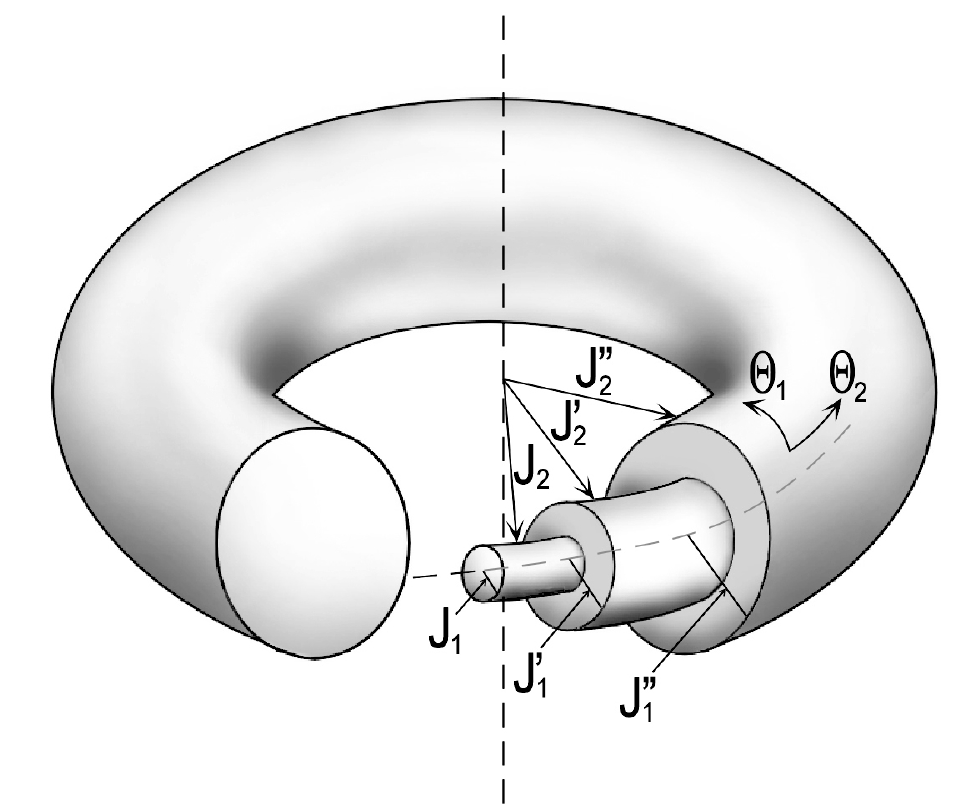} 
\caption{Nested tori for several values of the actions in two degrees of freedom. The vectors $\mathbf{J}=(J_1,J_2)$, 
$\mathbf{J'}=(J'_1,J'_2)$ and $\mathbf{J''}=(J''_1,J''_2)$ label the different tori and correspond to different initial conditions. $(\theta_1,\theta_2)$ are the angle coordinates on a given torus.}
\label{Fig3}
\end{center}
\end{figure}

We shall see below that the new coordinates, defined by 
\begin{equation}
\mbox{\boldmath$\theta$}=\frac{\partial S}{\partial\mathbf{J}},
\label{tc}
\end{equation}
play the role of angles, increasing by $2\pi$ when the phase-space trajectory undergoes a complete turn around $\Gamma_i$, while momenta (i.e. the actions) are the radii of the invariant tori (see Fig. \ref{Fig3}).

In action-angle variables, Hamilton equations read 
\begin{equation}
\mbox{\boldmath$\dot{\theta}$}=\frac{\partial{H}'}{\partial\mathbf{J}},\qquad
\mathbf{\dot{J}}=-\frac{\partial{H}'}{\partial \mbox{\boldmath$\theta$}},
\label{he_aa}
\end{equation}
where $H'=H$ is the transformed Hamiltonian which coincides with the original one since we are dealing with a time independent canonical transformation. 

However, $\mathbf{\dot{J}}=0$, because the $J$'s are constants of motion. Hence, $\partial{H}'/\partial\mbox{\boldmath$\theta$}=0$, and the transformed Hamiltonian only depends on the actions:
$$
{H}'={H}(\mathbf{J}),
$$
whence 
\begin{equation}
\mbox{\boldmath$\dot{\theta}$}=\frac{\partial{H}(\mathbf{J})}{\partial\mathbf{J}}\equiv\mbox{\boldmath$\omega$} \left( \mathbf{J} \right)=\mbox{constant}.
\label{omega}
\end{equation}
The solution of the Hamilton equations (\ref{he_aa}) thus reads 
\begin{equation}
J_i=\mbox{constant}, \qquad \theta_i(t)=\omega_i t+\beta_i,
\label{2-3}
\end{equation}
where $\beta_i=\theta_i(0)$ $\left( i=1,\dots,n \right)$. 

In action-angle variables, the equations of motion are, therefore, very simple: the actions (transformed momenta) are constants and the angle variables (transformed coordinates) are linear functions of time. In the original variables, $\left( \mathbf{q},\mathbf{p} \right)$, the equations of motion are obtained by introducing (\ref{2-3}) into the equations of the canonical transformation $(\mathbf{q},\mathbf{p}) \longrightarrow (\mbox{\boldmath$\theta$},\mathbf{J})$ which gives the original variables in terms of action-angle variables. 

Before proceeding further let us note that the transformation to action-angle variables is global. That is, the entire phase space is filled with tori (see Fig. \ref{Fig3}) and a given trajectory will forever reside on the surface of the particular torus fixed by the initial conditions $(\mathbf{q}(0),\mathbf{p}(0))$. Indeed, the choice of any invariant torus depends on the value of the constants $\alpha_i=I_i(\mathbf{q}(0),\mathbf{p}(0))$. The set of constants $\alpha_i$ determines the torus where the trajectory stays and the value of the angle variable $\theta_i$, at a given time, determines the position of the trajectory on that torus.

We will see next that {\it the $\omega$'s are the angular frequencies of the multiply periodic motion of the system}. To this end we first show that when the state of the system performs a complete oscillation along a closed curve $\Gamma_i$, the corresponding angle variable varies in $2\pi$. In effect, let us denote by $\Delta_i\theta_j$ the change of the angle variable $\theta_j$ when the coordinate $q_i$ realizes a complete oscillation (also known by the astronomical name of ``libration''). In the \ref{appenA} we show that 
\begin{equation}
\Delta_i\theta_j=2\pi\delta_{ij}.
\label{delta_ij}
\end{equation}
In other words, the angle variable $\theta_i$ changes in $2\pi$ only when the system executes a complete oscillation along the curve $\Gamma_i$ and no change otherwise. 

On the other hand, let $T_i$ ($i=1,\dots,n$) be the different periods of each libration motion. We see from  (\ref{2-3}) that the change in $\theta_i$ after a period $T_i$ is $$
\Delta_i\theta_i=\theta_i(t+T_i)-\theta_i(t)=\omega_i T_i,
$$
and equating it with (\ref{delta_ij}) we get
$$
T_i=\frac{2\pi}{\omega_i}.
$$
Therefore, the quantities $\omega_i$ --defined in (\ref{omega}) as derivatives of the Hamiltonian with respect to the actions-- are the angular frequencies of the (multiperiodic) motion of the system. This is quite remarkable since, for integrable Hamiltonians described by action-angle variables, to obtain the frequencies of the motion it is not necessary to integrate Hamilton equations, it suffices to take the derivative of the Hamiltonian with respect to the actions, a much easier task indeed.

\subsection{Periodic and multiply periodic motion} 
\label{subsec4-1}

Recall that the original variables, $\left( \mathbf{q},\mathbf{p} \right)$, and the action-angle variables, 
$\left( \mbox{\boldmath$\theta$},\mathbf{J} \right)$, are related by a canonical transformation with the generating function (\ref{S}). This obviously implies that the relation between these two sets of canonical variables is known. Let us denote this relation by
\begin{equation}
\mathbf{q}=\mathbf{q}(\mbox{\boldmath$\theta$},\mathbf{J}), \qquad \mathbf{p}=\mathbf{p}(\mbox{\boldmath$\theta$},\mathbf{J}).
\label{relation}
\end{equation}
Owing to the periodic nature of the angle variables, as expressed in equation (\ref{delta_ij}), we see from (\ref{relation}) that the $q$'s and $p$'s are multiply periodic functions of the $\theta$'s with periods $2\pi$, and the same is true for any dynamical function $f(\mbox{\boldmath$\theta$},\mathbf{J})$. 

As is well known from analysis, any multiply periodic function can be represented by a multiple Fourier series. Thus, for instance, $q_j$ would appear as
$$
q_j=\sum_{k_1=-\infty}^\infty \cdots \sum_{k_n=-\infty}^\infty c^{(j)}_{k_1,\dots,k_n}
\exp\{i(k_1\theta_1+\cdots+k_n\theta_n)\},
$$
where the $k_i$'s are integer indices running from $-\infty$ to $\infty$. By treating the set of $k_i$'s also as an $n$ dimensional vector, $\mathbf{k}=(k_1,\dots,k_n),$ the expansion can be written in a more compact form:
$$
q_j=\sum_{\mathbf{k}}c^{(j)}_\mathbf{k} e^{i\mathbf{k}\cdot\mbox{\boldmath$\theta$}},
$$
where we use the abbreviated notation:
$$
\sum_{\mathbf{k}}=\sum_{k_1=-\infty}^\infty\cdots\sum_{k_n=-\infty}^\infty.
$$
The Fourier coefficients $c^{(j)}_\mathbf{k}$, which depend on the actions, are given by
$$
c^{(j)}_\mathbf{k}(\mathbf{J})=\frac{1}{(2\pi)^n}\int_0^{2\pi}d\theta_1\cdots\int_0^{2\pi}d\theta_n\ e^{-i\mathbf{k}\cdot\mbox{\boldmath$\theta$}}q_j(\mbox{\boldmath$\theta$},\mathbf{J}).
$$

If we also write the time dependence of the angle variables --given in (\ref{2-3})-- as a vector equation,
$$
\mbox{\boldmath$\theta$}=\mbox{\boldmath$\omega$}t+\mbox{\boldmath$\beta$},
$$
then the time dependence of $q_j$ appears in the form
\begin{equation}
q_j(t)=\sum_{\mathbf{k}}a^{(j)}_\mathbf{k}(\mathbf{J}) e^{i\mathbf{k}\cdot\mbox{\boldmath$\omega$}t},
\label{q(t)}
\end{equation}
where $a^{(j)}_\mathbf{k}=c^{(j)}_\mathbf{k}e^{i\mathbf{k}\cdot\mbox{\boldmath$\beta$}}$. We easily convince ourselves that a similar expression holds for any momentum: 
\begin{equation}
p_j(t)=\sum_{\mathbf{k}}b^{(j)}_\mathbf{k}(\mathbf{J}) e^{i\mathbf{k}\cdot\mbox{\boldmath$\omega$}t}.
\label{p(t)}
\end{equation}
Notice that the right hand side of equations (\ref{q(t)}) and (\ref{p(t)}) correspond to a superposition of the equations of motion of $n$ linear oscillators. Therefore,

\begin{quotation}

\noindent
{\it the coordinates and momenta of integrable systems can be represented as a sum of n dimensional harmonic oscillators with angular frequencies given by the fundamental ones, $\mbox{\boldmath$\omega$}=(\omega_1,\dots,\omega_n),$ and their harmonics  $\mathbf{k}\cdot\mbox{\boldmath$\omega$}=k_1\omega_1+\cdots+k_n\omega_n,$ $(k_j=0,\pm 1,\pm 2,\cdots)$}.

\end{quotation}

\noindent
The coordinates and momenta of integrable systems are multiply periodic functions of time with $n$ independent (angular) frequencies $\omega_1,\dots,\omega_n$. However, this property {\it does not imply that, in general, the $q$'s and $p$'s are (simply) periodic functions of time}; for it would be necessary that there exist {\it a single period} $T_0$ for which all $q$'s and $p$'s are periodic: 
$$
\mathbf{q}(t+T_0)=\mathbf{q(t)}\qquad {\rm and} \qquad \mathbf{p}(t+T_0)=\mathbf{p}(t).
$$

In the \ref{appenB} we show that this is the case if, and only if, the frequencies $\omega_1,\dots,\omega_n$ are integer multiples of a single frequency $\omega_0$:
\begin{equation}
\omega_j=l_j\omega_0,
\label{fundamental}
\end{equation}
$(j=1,\dots,n)$, where $l_j=0,\pm 1,\pm 2,\cdots$ are integer numbers, and 
$$
\omega_0=\frac{2\pi}{T_0}.
$$ 
Equation (\ref{fundamental}) means that in order to have periodic motion, the frequencies must be {\it commensurable}. From (\ref{fundamental}) we immediately see that this is equivalent to assuming that all frequencies are rational multiples of each other
\begin{equation}
\frac{\omega_i}{\omega_j}=\frac{l_i}{l_j}=\mbox{a rational number}.
\label{commen_2}
\end{equation}
If the frequencies are {\it incommensurable}, in other words, if they are not rationally related, then the motion is termed multiply periodic or quasiperiodic or conditionally periodic, according to different terminologies in use, and never repeats itself. 

\begin{figure}
\begin{center}
\includegraphics[scale=0.70]{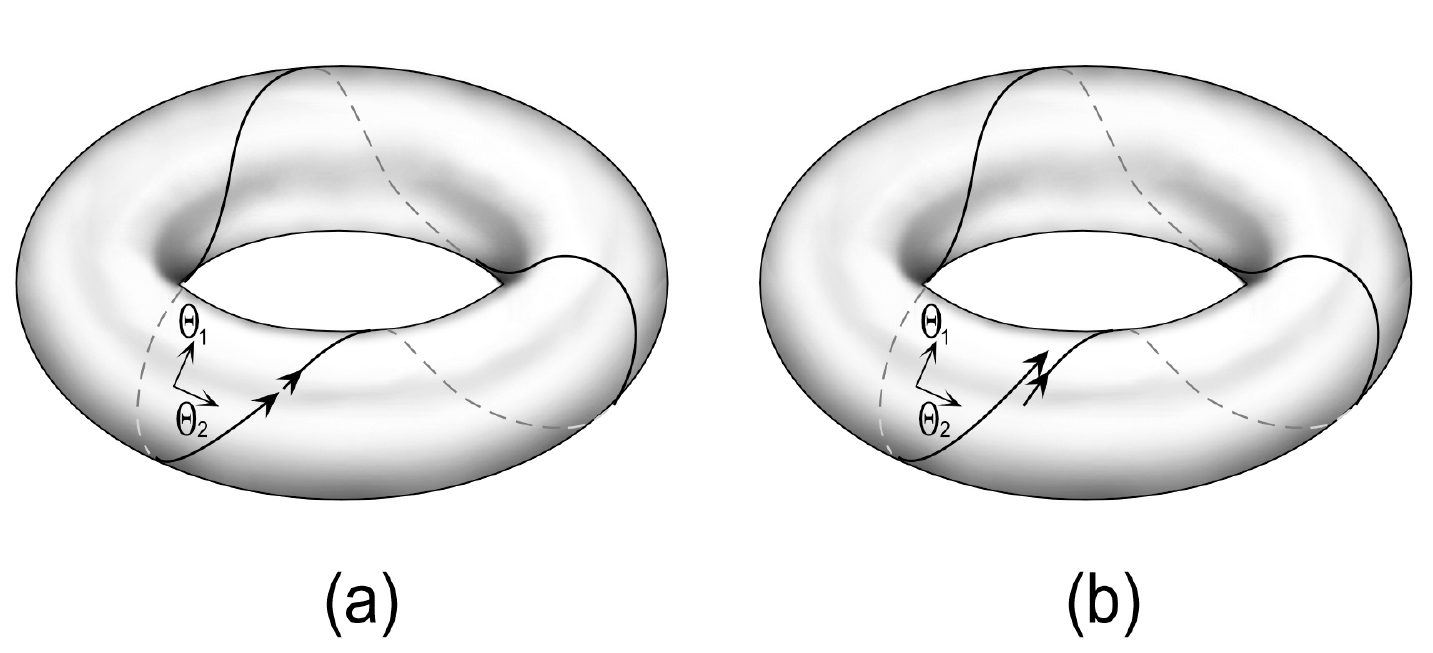} 
\caption{Comparison of phase-space trajectories on $2$ dimensional tori. The torus in (a) is a rational surface (resonant torus) where $\omega_1/\omega_2=3$, that is, the trajectory closes over itself after three turns around $\Gamma_1$ and one turn around $\Gamma_2$. The torus in (b) is an irrational surface (nonresonant torus) where the frequencies are not commensurable. In (b) the phase-space trajectory will eventually cover the surface of the torus densely.}
\label{Fig4}
\end{center}
\end{figure}

A formal condition for the frequencies to be commensurable is the existence of $n-1$ relations of the form
\begin{equation}
\sum_{j=1}^nl_{k_{j}}\omega_j=0, \quad\qquad (k=1,\dots,n-1),
\label{commen_3}
\end{equation}
where $l_{k_{j}}=0\pm 1\pm 2\pm\cdots$ are integers. Indeed, if (\ref{commen_3}) holds, then we have an homogeneous system of $n-1$ algebraic equations (with integer coefficients) for $n$ unknowns, the frequencies, which means that we can solve (\ref{commen_3}) by writing all frequencies in terms of one of them, say $\omega_1$. In other words, we obtain $\omega_j=(m_j/m_1)\omega_1$, where $m_1$ and $m_j$ are integers. Therefore,  $\omega_j/\omega_i=m_j/m_i$ is a rational number, and all frequencies are commensurable.  

Let us remember that the phase-space trajectories lie on the invariant tori defined by equation (\ref{2-3}), and since these curves are mathematically described by functions of coordinates and momenta, the above considerations on single or multiple periodicity apply to phase-space trajectories as well. We can, therefore, conclude that {\it on a given torus, the phase-space trajectory will be a closed curve (i.e. the motion of the system will be periodic) if and only if the frequencies of the motion are commensurable. When frequencies are incommensurable the trajectory will densely cover the torus, never closing on itself}. In the first case, we call it {\it rational, or resonant, torus}, while in the latter {\it irrational, or nonresonant, torus} (Fig. \ref{Fig4}).

\subsection{The Kepler problem}
\label{sec4-2}

A classic example of an integrable Hamiltonian is provided by the Kepler problem. Consider the relative motion of a planet of mass $m$ orbiting around a star of mass $M$. The Hamiltonian of the system in polar coordinates, $\left( r,\phi \right)$, is
\begin{equation}
H_0=\frac{p_r^2}{2\mu}+\frac{p_{\phi}^2}{2\mu r^2}-
\frac{k}{r},
\label{2-4}
\end{equation}
where $\left( p_r,p_{\phi} \right)$ is the relative momentum of the two bodies, $\mu=mM/\left( m+M \right)$ is the reduced mass and $k=GmM$ ($G$ is the gravitational constant).

Both the energy, $H_0=E$, and the angular momentum, $\mathbf{L}$, of the two-body system are constants of the motion. The latter implies that the motion takes place on the plane perpendicular to the direction of $\mathbf{L}$. 

After a canonical transformation from the original variables $\left( p_r,p_{\phi},r,\phi \right)$ to the action-angle variables $\left( J_1,J_2,\theta_1,\theta_2 \right)$, the Hamiltonian is only a function of the actions \cite{born,goldstein,landau}:
\begin{equation}
H_0=\frac{-\mu k^2}{2 \left( J_1+J_2 \right)^2}.
\label{2-5}
\end{equation}

The equations of motion in the original variables are quite complicated (elliptic orbits), but in terms of action-angle variables they are very simple:
\begin{equation}
J_i=\alpha_i,\qquad\theta_i \left( t \right)=\omega_it+\beta_i,
\label{2-6}
\end{equation}
$(i=1,2)$ where $\alpha_i$ and $\beta_i$ are constants determined by the initial conditions. 

The motion of the system in the phase space (i.e., the phase-space trajectory) lies on a two dimensional torus embedded in the three dimensional manifold of constant energy. The torus has two angular coordinates $\theta_1$ and $\theta_2$ and two constant radii $J_1=\alpha_1$ and $J_2=\alpha_2$ which are fixed by the initial conditions. 

There are two frequencies, $\omega_1$ and $\omega_2$, associated to the motion; If they are commensurable, that is if $l_1\omega_1=l_2\omega_2$, where $l_1$ and $l_2$ are integers, the trajectory is closed and the motion is periodic (the trajectory closes on itself after $l_1$ turns around $\Gamma_1$ and $l_2$ turns around $\Gamma_2$). If the frequencies are incommensurable the trajectory will never repeat itself and will eventually cover the entire surface of the torus. 

For the Kepler problem the frequencies are (recall that $\omega_i=\dot{\theta}_i=\partial H_0/\partial J_i$) 
$$
\omega_1=\omega_2=\frac{1}{k}\sqrt{-\frac{2H_0^3}{\mu}},
$$
which is Kepler's third law. The frequencies are commensurable (in fact, equal) and hence the phase-space trajectory is always closed. In other words, the motion is periodic.

\section{Nonintegrable Hamiltonians.} 

\begin{figure}
\begin{center}
\includegraphics[scale=0.70]{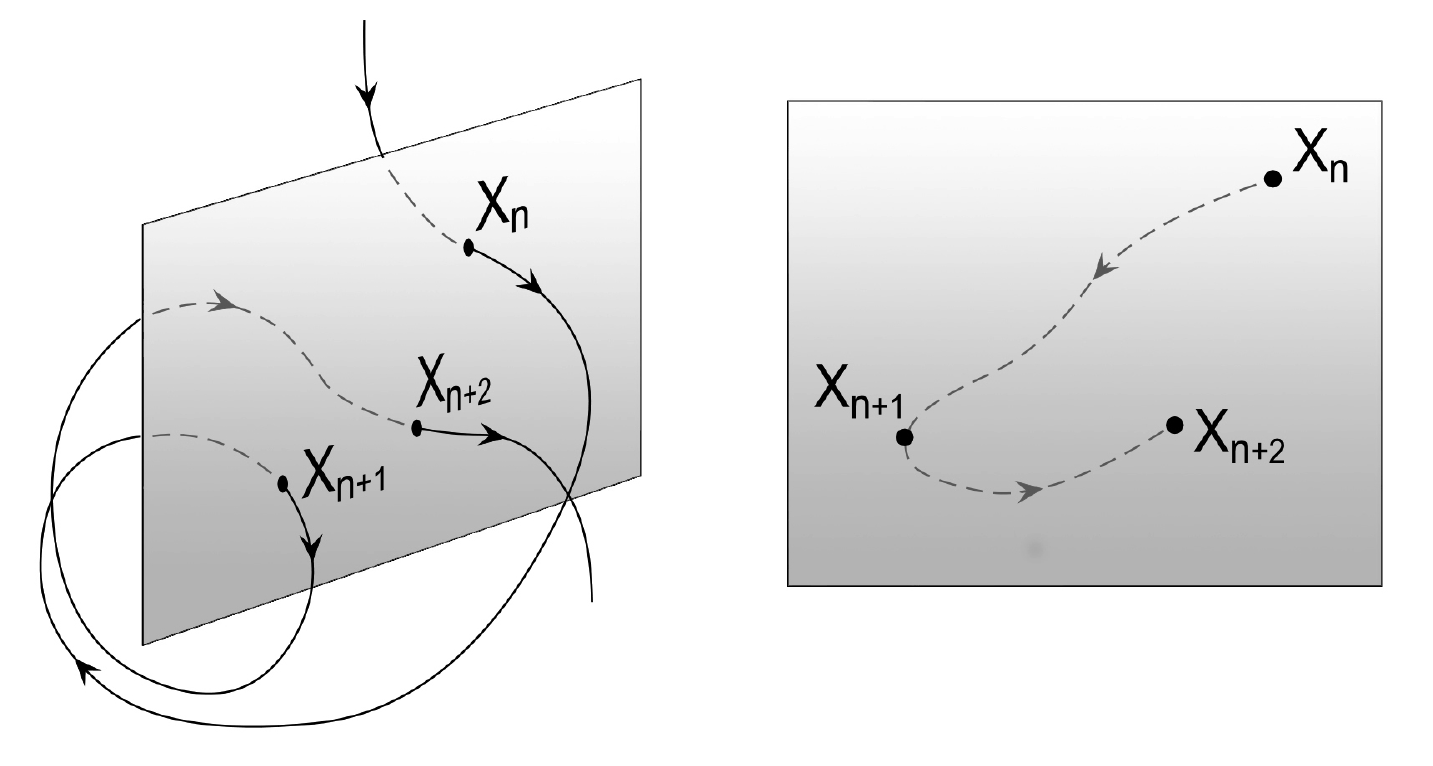} 
\caption{Intersections of a phase-space trajectory with a Poincar\'e surface of section.}
\label{Fig5}
\end{center}
\end{figure}

We know that the existence of as many constants of motion in involution as degrees of freedom determines integrability and regular motion.  On the other hand, conservative systems with one degree of freedom always have one integral of motion, the energy, and they are, therefore, integrable. The following question then suggests itself: can we know whether or not a conservative system with $n>1$ degrees of freedom is integrable? Unfortunately there is no general answer to this problem. In most situations, integrability is guessed, but not proved, by constructing a \textit{Poincar{\'e} surface of section} which we can define as any bidimensional surface, contained in the phase space, where one can conveniently analyze the successive intersections of the phase-space trajectory (Fig. \ref{Fig5}).

To see how this concept works, let us suppose a conservative system with two degrees of freedom. The Hamiltonian is a constant of motion and we write:
\begin{equation}
H(q_1,q_2; p_1,p_2)=E.
\label{4-1}
\end{equation}
This restricts  phase-space trajectories to evolve inside a $3$ dimensional manifold --the surface of constant energy or ``energy shell'' represented by (\ref{4-1})--  embedded in the phase space which has dimension $4$. Suppose the system has a second integral (for instance, the angular momentum):
\begin{equation}
I(q_1,q_2; p_1,p_2)=\alpha,
\label{4-2}
\end{equation}
where $\alpha$ is an arbitrary constant; then this integral also defines another $3$ dimensional manifold in the $4$ dimensional phase space and any trajectory must also evolve in it.

The coordinates and momenta of the system cannot take arbitrary values because they must satisfy the conservation laws (\ref{4-1}) and (\ref{4-2}). From a geometrical point of view, this means that any phase-space trajectory is bounded to move on a $2$ dimensional surface which is the intersection of two $3$ dimensional manifolds $H=E$ and $I=\alpha$. Indeed, from the conservation of energy (\ref{4-1}), we can write, for instance, $p_2$ as a function of $p_1, q_1, q_2$ and the energy $E$, that is, $p_2=p_2(q_1,q_2,p_1;E)$. Substituting this expression into the second conservation law (\ref{4-2}) we can obtain $p_1$ in terms of $q_1$ and $q_2$ and the constants $E$ and $\alpha$:  
\begin{equation}
p_1=\phi \left( q_1,q_2;E,\alpha \right).
\label{4-3}
\end{equation}
This is the equation of a $2$ dimensional surface within the $4$ dimensional phase space. Such a surface is determined by the specific values of the constants  $E$ and $\alpha$, which, in turn, depend on the initial conditions. 

\begin{figure}
\begin{center}
\includegraphics[scale=0.60]{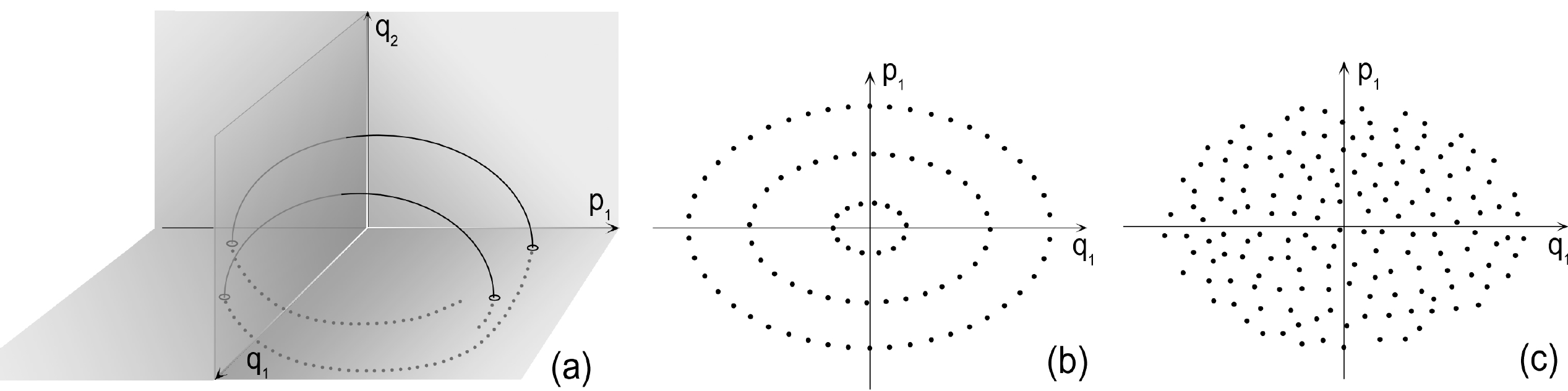} 
\caption{Construction of a Poincar\'e surface of section for a system with two degrees of freedom. (a) We can obtain a Poincar\'e section plotting a point each time $q_2=0$, i.e., each time the trajectory passes through the plane $(q_1,p_1)$. (b) If there are two integrals of motion in involution, the trajectory will remain along one dimensional curves, $p_1=\phi(q_1,0;E,\alpha)$, in the plane $(q_1,p_1)$. (c) If only one constant of motion exists (the energy), the trajectory will disperse over a region of the $(q_1,p_1)$ plane.}
\label{Fig6}
\end{center}
\end{figure}

If we now consider the section $q_2=0$ of the surface (\ref{4-3}) --and this is an example of a Poincar{\'e} surface of section-- we see that the successive intersections of the phase-space trajectory with this Poincar{\'e} section are along the curve 
$$
p_1=\phi \left( q_1,0;E,\alpha \right) 
$$
on the plane $(q_1,p_1)$.

In general, for a given conservative Hamiltonian, we do not know whether or not a second integral of motion exists. We can guess, nonetheless, its existence by solving Hamilton equations numerically and then drawing $p_1$ and $q_1$ each time $q_2=0$ (Fig. \ref{Fig6} (a)). If the system is integrable, the intersections of the trajectory in the Poincar{\'e} section $q_2=0$ will appear as a collection of points regularly distributed on the curve $p_1=\phi \left( q_1,0;E,\alpha \right)$ (Fig. \ref{Fig6} (b)). On the other hand, if the system is not integrable the intersections of the phase-space trajectory will appear randomly --that is, chaotically-- scattered over a region only limited by the conservation of energy (Fig. \ref{Fig6} (c)).

This was the method used by Henon and Heiles \cite{henon} to determine the possible existence of a third integral which, as the observations showed, constrained the motion of a star in a galaxy with an axis of symmetry. This system has three degrees of freedom and two known constants of motion, the energy and the component of the angular momentum along the direction of the symmetry axis. For a long time it was believed that there was no third integral of motion because it had not been found analytically. 

\begin{figure}
\begin{center}
\includegraphics[scale=0.70]{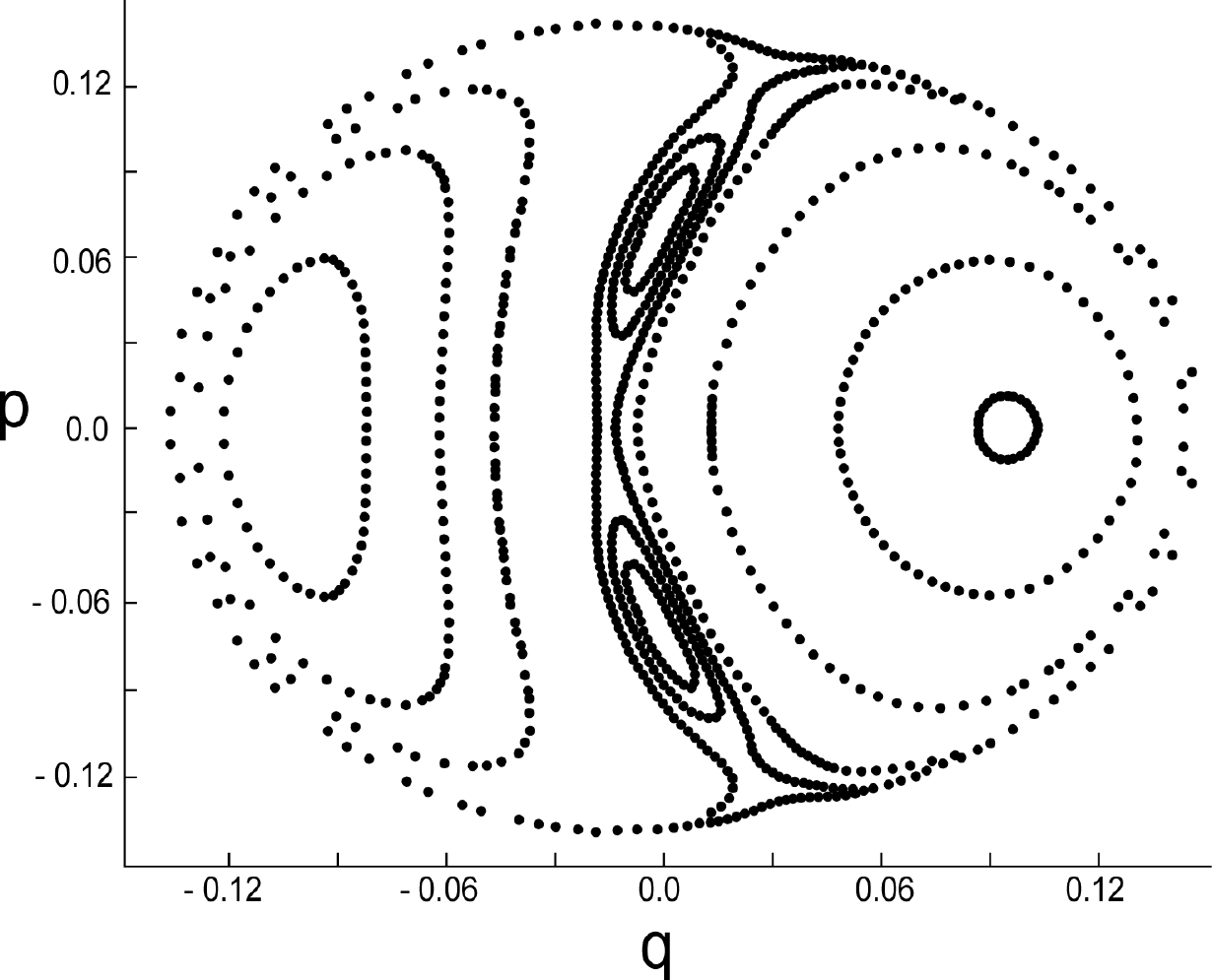} 
\caption{Poincar\'e section for the Henon-Heiles Hamiltonian at low energy $E=0.01$.}
\label{Fig7}
\end{center}
\end{figure}

\begin{figure}
\begin{center}
\includegraphics[scale=0.70]{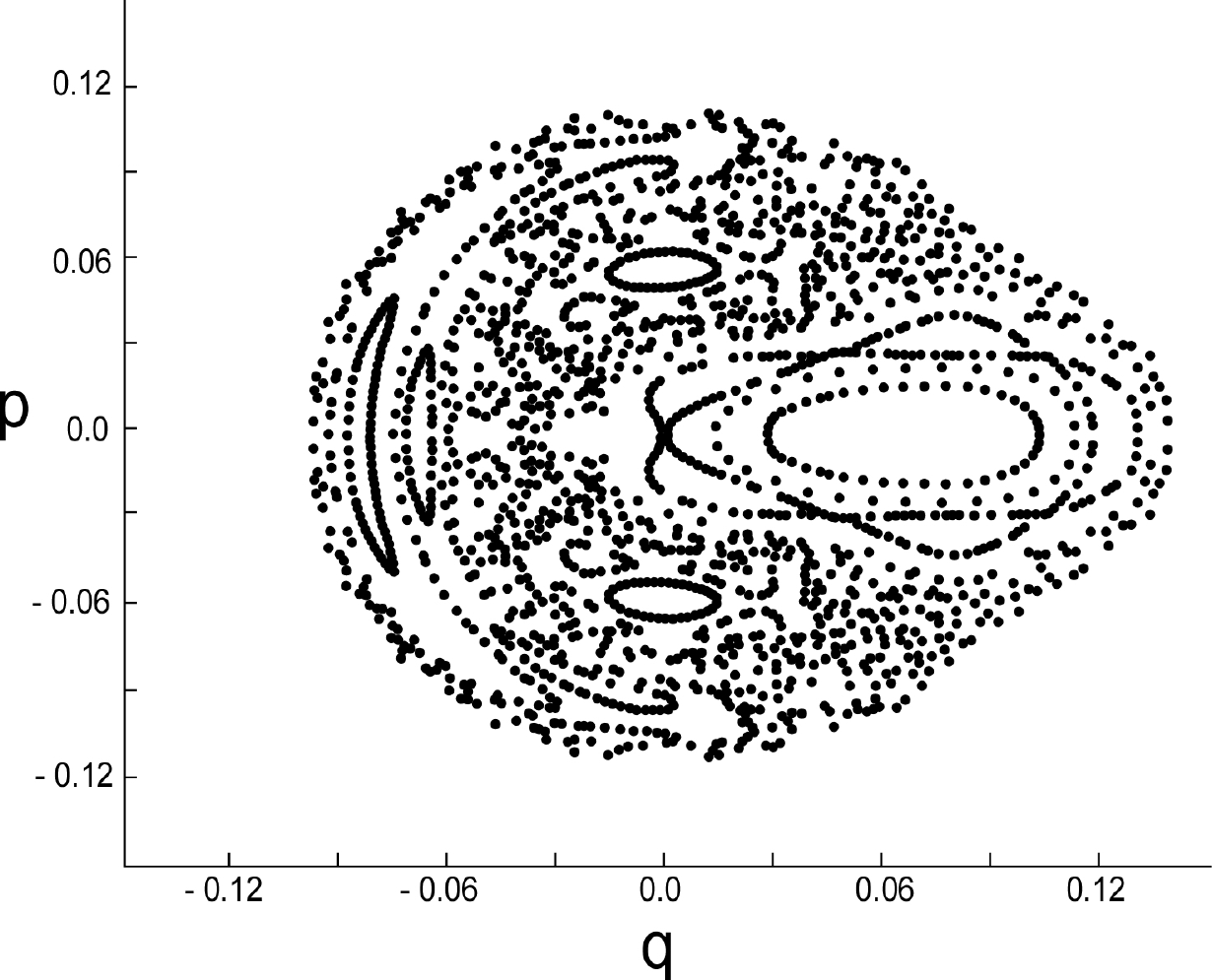} 
\caption{Poincar\'e section for the Henon-Heiles Hamiltonian at high energy $E=0.12$.}
\label{Fig8}
\end{center}
\end{figure}

Henon and Heiles proposed the following Hamiltonian --without any known symmetries leading to a third integral-- that could model the basic traits of the problem \cite{henon,lichtenberg}:
$$
H=\frac{1}{2}\left(p_1^2+p_2^2\right)+
\frac{1}{2}\left(q_1^2+q_2^2+2q_1^2q_2-\frac{2}{3}q_2^3\right)=E,
$$
and numerically studied the behaviour of the equations of motion. The Hamilton equations are
$$
\dot{q_i}=p_i,\qquad \dot{p}_1=-q_1-2q_1q_2,\qquad\dot{p}_2=-q_2-2q_1^2-2q_2^2,
$$
$\left( i=1,2 \right)$. Observe the occurrence  of non-linear quadratic terms in these equations of motion which are consequence of the cubic terms in the potential energy. Note also that the magnitude of these non-linear terms increases with the energy. The representation of the intersections of the phase-space trajectory on the Poincar{\'e} section $q_2=0$ is given in Figs. \ref{Fig7}-\ref{Fig8}. For low energies (Fig. \ref{Fig7}), the distribution of points is apparently regular, which seems to suggest the existence of a third integral, at least to the accuracy of the plotting. As the energy increases (which increases the effect of the nonlinear terms) the third integral appears to be destroyed, at least partially, but still remain some ``islands''  where the motion is regular (Fig. \ref{Fig8}). For even higher energies the hypothetical third integral is completely destroyed. The scattered points appearing in Fig. \ref{Fig8} correspond to a single phase-space trajectory that, for obvious reasons, we will call chaotic, and which implies a highly sensitive dependence on initial conditions that makes the long-time evolution of the system totally unpredictable.

\section{Canonical perturbation theory. Resonances}

As stated in previous sections, most mechanical systems are not integrable, having only the energy as a constant of motion. However, in many situations, integrable and nonintegrable Hamiltonians  differ very little. In such cases we can obtain approximate solutions of the equations of motion by means of the canonical perturbation theory. We will outline this theory because it provides the clue to the appearance of chaos in deterministic systems, for it shows that chaotic behaviour is a direct consequence of the occurrence of resonances, i.e., of small denominators leading to infinities. 

In order to fix ideas, let us return to the Kepler problem discussed in section \ref{sec4-2} and suppose that, in addition to the planet orbiting the Sun, there is another celestial body which perturbs the motion of the two-body system. We assume that the effect of this perturbation is small (because, for instance, the third body is small or more distant), so that the total Hamiltonian can be written in the form
\begin{equation}
H=H_0 \left( J_1,J_2 \right)+\epsilon H_1 \left( J_1,J_2,\theta_1,\theta_2 \right),
\label{5-1}
\end{equation}
where $\epsilon\ll 1$ is a small parameter and $H_0$ is the unperturbed Hamiltonian of the Kepler problem as given in (\ref{2-5}). 

According to our previous discussions, we know that the unperturbed motion, represented by $H_0$, takes place on a bidimensional torus of radii $J_i=\alpha_i$, $\left( i=1,2 \right)$. We want now to find the corrections to the unperturbed motion due to the presence of a third body. As we generally cannot obtain an exact solution to the equations of motion, we search for approximations via perturbation expansions in powers of $\epsilon$. 

Observe that the motion generated by the Hamiltonian (\ref{5-1}) is bounded. This allows us to expand the perturbation as a Fourier series in the angular variables. Therefore, following the notation of section \ref{subsec4-1}, we see that (\ref{5-1}) can be written as
\begin{equation}
H(\mbox{\boldmath$\theta$},\mathbf{J})=H_0 \left(\mathbf{J} \right)+ \epsilon\sum_{\mathbf{k}}H^{(1)}_{\mathbf{k}}(\mathbf{J})\cos\left(\mathbf{k}\cdot\mbox{\boldmath$\theta$}\right),
\label{5-2}
\end{equation}
where $\mbox{\boldmath$\theta$}=(\theta_1,\dots,\theta_n),$ $\mathbf{J}=(J_1,\dots,J_n)$ and $\mathbf{k}=(k_1,\dots,k_n)$ ($k_i=0,\pm 1,\pm 2,\cdots$). The Fourier coefficients, $H^{(1)}_{\mathbf{k}}(\mathbf{J})$, are defined by
\begin{equation}
H^{(1)}_{\mathbf{k}}(\mathbf{J})=
\frac{1}{(2\pi)^n}\int_0^{2\pi}d\theta_1\cdots\int_0^{2\pi}d\theta_n\  H_1(\mathbf{J},\mbox{\boldmath$\theta$})\cos\left(\mathbf{k}\cdot\mbox{\boldmath$\theta$}\right).
\label{FC}
\end{equation}
Let us note that we have generalized (\ref{5-1}) to include multiply periodic systems with $n$ degrees of freedom.

The main idea behind canonical perturbation theory is simple and consists in finding, by means of a power series in $\epsilon$, a canonical transformation,
\begin{equation}
(\mbox{\boldmath$\theta$}, \mathbf{J}) \longrightarrow (\mbox{\boldmath$\theta'$}, \mathbf{J'}),
\label{CT}
\end{equation}
to a new set of action-angle variables such that {\it the transformed Hamiltonian ${H}'(\mathbf{J'})$ is only a function of the new actions $\mathbf{J'}$}. If we could achieve this, then the Hamiltonian (\ref{5-1}) would become completely integrable  and its equations of motion would be easily obtained; that is, the $J'$'s would be constants and the $\theta'$'s linear functions of time. 

Unfortunately, this procedure cannot be done exactly. However, we may proceed by perturbation expansions since $\epsilon$ is small. We  take this path and introduce the following generating function for the canonical transformation (\ref{CT}):
\begin{equation}
S(\mbox{\boldmath$\theta$}, \mathbf{J'})=\mbox{\boldmath$\theta$}\cdot\mathbf{J'}
+\epsilon\sum_{\mathbf{k}} S_{\mathbf{k}}(\mathbf{J'})\sin\left(\mathbf{k}\cdot\mbox{\boldmath$\theta$}\right)+{\rm O}(\epsilon^2),
\label{5-3}
\end{equation}
which is defined by a perturbation series up to first order in $\epsilon$ with the unknown Fourier coefficients $S_{\mathbf{k}}(\mathbf{J'})$ to be determined later. In terms of the generating function, the canonical transformation is obtained by (cf (\ref{canonical_3})) 
$$
\mathbf{J}=\frac{\partial S}{\partial\mbox{\boldmath$\theta$}}, \qquad \mbox{\boldmath$\theta'$}=\frac{\partial S}{\partial\mathbf{J'}}.
$$
That is
\begin{equation}
\mathbf{J}=\mathbf{J'}+\epsilon\sum_{\mathbf{k}} \mathbf{k} S_{\mathbf{k}}(\mathbf{J'})\cos\left(\mathbf{k}\cdot\mbox{\boldmath$\theta$}\right)+{\rm O}(\epsilon^2),
\label{J}
\end{equation}
\begin{equation}
\mbox{\boldmath$\theta'$}=\mbox{\boldmath$\theta$}+
\epsilon \sum_{\mathbf{k}} \frac{\partial S_{\mathbf{k}}(\mathbf{J'})}{\partial\mathbf{J'}}\sin\left(\mathbf{k}\cdot\mbox{\boldmath$\theta$}\right)
+{\rm O}(\epsilon^2).
\label{theta}
\end{equation}

Since we are dealing with a time-independent canonical transformation, the transformed Hamiltonian, $H'=H$, coincides with the original one. Therefore, the new Hamiltonian is found from (\ref{5-2}) by solving the equations (\ref{J}) and (\ref{theta}) for $(\mbox{\boldmath$\theta$}, \mathbf{J})$ as a function of $(\mbox{\boldmath$\theta'$}, \mathbf{J'})$. We detail this procedure in the \ref{appenC} where we show that 
\begin{eqnarray}
H'(\mbox{\boldmath$\theta'$}, \mathbf{J'})&=&H_0(\mathbf{J'})+\epsilon H_{\mathbf{0}}^{(1)}(\mathbf{J'}) \nonumber \\
&+&\epsilon\sum_{\mathbf{k}\neq\mathbf{0}}\left[H^{(1)}_{\mathbf{k}}(\mathbf{J'})+\mathbf{k}\cdot\mbox{\boldmath$\omega$}
S_{\mathbf{k}}(\mathbf{J'})\right]\cos\left(\mathbf{k}\cdot\mbox{\boldmath$\theta$}\right)+{\rm O}(\epsilon^2).
\label{H'}
\end{eqnarray}
In this expressions $\mbox{\boldmath$\omega$}=\mbox{\boldmath$\omega$}(\mathbf{J'})$ are the (known) frequencies of the unperturbed motion:
\begin{equation}
\mbox{\boldmath$\omega$}=\frac{\partial H_0}{\partial \mathbf{J'}},
\label{unperturb_freq}
\end{equation}
whereas $H_{\mathbf{0}}^{(1)}(\mathbf{J'})$ is the Fourier coefficient of the perturbation corresponding to $\mathbf{k}=\mathbf{0}$ and reads (see (\ref{FC}))
\begin{equation}
H_{\mathbf{0}}^{(1)}(\mathbf{J'})=\frac{1}{(2\pi)^n}\int_0^{2\pi}d\theta_1\cdots\int_0^{2\pi}d\theta_n\ H_1(\mathbf{J'},\mbox{\boldmath$\theta$}).
\label{5-6a}
\end{equation}
In other words, $H_{\mathbf{0}}^{(1)}$ {\it is the average of the perturbation $H_1$ with respect to the original angle variables}. 

Recall that the Fourier coefficients $S_{\mathbf{k}}(\mathbf{J'})$ of the generating function (\ref{5-3}) are still unknown and we are, therefore, free to choose them in the most convenient way. Let us take
\begin{equation}
S_{\mathbf{k}}(\mathbf{J'})=-\frac{H^{(1)}_{\mathbf{k}}(\mathbf{J'})}{\mathbf{k}\cdot\mbox{\boldmath$\omega$}},
\label{5-7}
\end{equation}
($\mathbf{k}\neq\mathbf{0}$), so that the term inside the sum in equation (\ref{H'}) vanishes; we get
\begin{equation}
H'=H_0(\mathbf{J'})+\epsilon H_{\mathbf{0}}^{(1)}(\mathbf{J'})+{ \rm O}\left(\epsilon^2 \right).
\label{5-8}
\end{equation}
In such a case, and to the first order in $\epsilon$, {\it the transformed Hamiltonian $H'$ is only a function of the new action variables and is, therefore, integrable (up to first order)}. 

Moreover, the perturbed frequencies $\mbox{\boldmath$\omega'$}$ (given by the derivatives of the transformed Hamiltonian $H'$ with respect to the actions) read 
\begin{equation}
\mbox{\boldmath$\omega'$}=\mbox{\boldmath$\omega$}+\epsilon\frac{\partial}{\partial\mathbf{J'}} H^{(1)}_{\mathbf{0}}(\mathbf{J'})+O(\epsilon^2),
\label{perturbed_freq}
\end{equation}
where $ H^{(1)}_{\mathbf{0}}(\mathbf{J'})$ is the average of the perturbation $H_1$ (cf (\ref{5-7})) and $\mbox{\boldmath$\omega$}$ are the unperturbed frequencies given by (\ref{unperturb_freq}).

Finally, the relation between the original actions and the new variables, is given by (see (\ref{J}) and (\ref{theta}))
\begin{equation}
\mathbf{J}= \mathbf{J'}-\epsilon\sum_{\mathbf{k}\neq\mathbf{0}} \mathbf{k}\frac{H^{(1)}_{\mathbf{k}}(\mathbf{J'})}{\mathbf{k}\cdot\mbox{\boldmath$\omega$}}\cos\left(\mathbf{k}\cdot\mbox{\boldmath$\theta'$}\right)+
{\rm O} \left( \epsilon^2 \right).
\label{5-9a}
\end{equation}
Equation (\ref{5-9a}) is, to the lowest order in $\epsilon$, the solution to our problem. We have obtained the new action variables $\mathbf{J'}$ which contain the first-order correction due to the perturbation. To first order in $\epsilon$ the $\mathbf{J'}$ are constants and the new angles $\mbox{\boldmath$\theta'$}$ linear functions of time.

Unfortunately, the expressions (\ref{5-7}) and (\ref{5-9a}) suffer a serious illness, for both $S_{\mathbf{k}}$ and $\mathbf{J}$ diverge if 
\begin{equation}
\mathbf{k}\cdot\mbox{\boldmath$\omega$}=k_1\omega_1+\cdots+k_n\omega_n=0.
\label{resonance}
\end{equation}
In other words, {\it the perturbation method breaks down when internal resonances occur inducing perturbation expansions to diverge}.

Let us observe that if the unperturbed motion takes place on a rational (resonant) torus, i.e., if the frequencies of the unperturbed Hamiltonian $H_0$ are commensurable, there always exists a set of integers such that condition (\ref{resonance}) holds (see (\ref{commen_3})). At first sight, this would imply the restriction of canonical perturbations to the motion on irrational (nonresonant) tori. However, the problem is worse, because, even if the frequencies are incommensurable (and recalling that any irrational number can be approximated, within any degree of accuracy, by a sequence of rational numbers) there will always be integers $k_1,\dots, k_n$ that generate denominators $\mathbf{k}\cdot\mbox{\boldmath$\omega$}$ as small as desired, with the result of convergence failure.\footnote{In the perturbed Kepler problem, as given in (\ref{5-1}), the term $k_1\omega_1+k_2\omega_2$ which appears in the denominator of the perturbation series, albeit never zero, can be arbitrarily small for $k_1$ and $k_2$ covering all integer numbers except $0$. Thus, for instance, Jupiter and Saturn, in their motion around the Sun cover $299$ and $120,5$ seconds of arc in a day and the denominator $5\omega_{Sat}-2\omega_{Jup}$ is very small compared with each frequency.} Hence we get two series whose convergence is uncertain. For one hand, we have Fourier series that cannot converge due to internal resonances which lead to the famous problem of ``small denominators''. On the other hand, far from resonant frequencies and also because of small divisors, it is quite unclear that the perturbation series in $\epsilon$ converge. In fact, Poincar{\'e} proved that all these series do not lead to a convergent representation of the perturbed problem \cite{whittaker}. Consequently, fundamental questions, as the long-time stability of planetary orbits in the solar system, could not be answered in a satisfactory way. Let us remind Moser's query and ask again: Is the solar system stable?

\section{The KAM Theorem. Stability and Chaos}

At the end of the 19th century, the general consensus was that the addition of even the smallest perturbation would render any conservative system nonintegrable, and that any phase-space trajectory would densely explore the energy shell in a chaotic manner. \footnote{When this occurs the system is termed ``ergodic'', a key concept in statistical mechanics because it implies that time averages are equivalent to phase-space averages.} This was the  first crack in classical certainty and a severe blow on classical thinking.

All efforts and attempts by some of the finest mathematicians and physicists of the time failed to solve the small divisor problem. Poincar{\'e} called it  the ``fundamental problem of classical mechanics'' and seemed to be an unsurmountable obstacle. Thus, around the beginning of the 20th century, this field of research dropped off and questions concerning long-time planetary stability were almost forgotten in mainstream research.

The breakthrough came in 1954, when A. N. Kolmogorov \cite{kolmogorov} devised a way of obtaining a theory of canonical perturbations which is rapidly convergent and appropriate to nonresonant tori. Kolmogorov's new idea was rigorously proved by V. I. Arnold in 1961 \cite{arnold(b)} and,  independently, by J. K. Moser in 1962 \cite{moser}. The complete result is known as the Kolmogorov-Arnold-Moser (KAM) Theorem.

Following Arnold (\cite{arnold(b),arnold}, see also \cite{tabor}) suppose that an integrable Hamiltonian $H_0$ is perturbed by a function $H_1$, so that the total Hamiltonian is
\begin{equation}
H=H_0 \left( \mathbf{J} \right)+\epsilon H_1 \left( \mathbf{J},
\mbox{\boldmath$\theta$} \right), 
\label{6-1}
\end{equation}
$(\epsilon\ll 1)$ where the perturbation $H_1$ is assumed to be periodic in the angular variables:
$$
H_1 \left( \mathbf{J},\theta_1+2\pi,\dots,\theta_n+2\pi \right)=H_1 \left( \mathbf{J},\theta_1,\dots,\theta_n \right).
$$
Hamilton equations read
\begin{equation}
\dot{J}_i=-\epsilon\frac{\partial H_1}{\partial\theta_i},\qquad
\dot{\theta}_i=\omega_i(\mbox{\boldmath{J}})+\epsilon
\frac{\partial H_1}{\partial J_i},
\label{6-2}
\end{equation}
$(i=1,2,\dots,n)$ where $\omega_i=\partial H_0/\partial J_i$ are the frequencies of the unperturbed motion. 

Kolmogorov's idea was that, far from resonant frequencies, it is possible to construct a perturbation series {\it in even powers of} $\epsilon$. This leads to a fast convergent (``superconvergent'') perturbation series with the overall result that, for most initial conditions, the motion generated by the Hamiltonian (\ref{6-1}) remains  quasiperiodic, that is, confined to reside on certain surfaces which resemble slightly deformed tori; while the complementary of this regular motion --that is, the chaotic motion-- has a small Lebesgue measure if $\epsilon$ is small. In other words, the set of chaotic trajectories is negligible as $\epsilon\rightarrow 0$.

In order to be more precise in the formulation of the theorem, one has to assume that $H$ is an analytic function and that the unperturbed motion generated by $H_0$ is nondegenerate. The last statement implies that the Hessian of $H_0$ is different from zero:
$$
\det\left|\frac{\partial^2H_0}{\partial J_i\partial J_j}\right|\neq 0.
$$
The next step consists in identifying, for the unperturbed system, a particular torus, $T_0 \left( \mbox{\boldmath$\omega^*$} \right)$, defined by a set of frequencies, $\mbox{\boldmath$\omega^*$}=(\omega_1^*,\dots,\omega_n^*)$, which are incommensurable, that is 
$$
k_1\omega_1^*+\cdots+k_n\omega_n^*\neq 0
$$
for all the integers $k_i\neq 0$. The KAM theorem can then be stated as \cite{arnold(b),arnold,tabor}:

\begin{quotation}
\noindent
{\it If $\epsilon H_1$ is small enough, then for almost all nonresonant frequencies $\mbox{\boldmath$\omega^*$}$ there exists an invariant surface $T\left(\mbox{\boldmath$\omega^*$}\right)$ of the perturbed system such that $T \left( \mbox{\boldmath$\omega^*$} \right)$ is close to $T_0 (\mbox{\boldmath$\omega^*$})$.}
\end{quotation}

\noindent
In other words, the trajectories of the perturbed Hamiltonian will evolve on surfaces $T^*=T \left( \mbox{\boldmath$\omega^*$} \right)$ which are ``close to'' tori $T_0$ of the integrable system. This means that, for almost all (but {\it not} all) nonresonant frequencies, the phase-space trajectories are regular and not chaotic as far as $\epsilon$ is small enough (the theorem does not tell how small $\epsilon$ has to be). 

Moreover, the theorem also proves that the measure of the set of points in phase space which do not lie on any surface $T^*$ tends to zero when $\epsilon\rightarrow 0$. That is to say, the probability of randomly choosing initial conditions not leading to motion on stable surfaces  $T^*$ is vanishingly small as $\epsilon\rightarrow 0$. This has momentous consequences for the stability of planetary orbits because it is extremely unlikely (though not impossible) that a given celestial body is located in some unstable (i.e., resonant) region (see, nonetheless, the next section). 

Therefore, we see that, under the conditions of the theorem, small perturbations do not destroy invariant tori $T_0$ but only deform them . On these deformed tori, called 
\textit{KAM surfaces}, the motion is regular and not chaotic.

The rigorous proof of the theorem is highly technical and we will not give it here. There are, however, more accessible derivations such, for instance, that of R. Barrar \cite{barrar} to which we refer the interested reader (see also \cite{reichl}). As mentioned above the clue of the proof, discovered by Kolmogorov, is a superconvergent perturbation method similar to the old Newton-Raphson technique for solving algebraic equations \cite{tabor}. 

KAM tori $T^*$ are determined by nonresonant frequencies $\mbox{\boldmath$\omega^*$}$ and on these deformed tori the motion is not chaotic. But, what about resonant tori? What happens to them? In other words, what is the fate, when a perturbation is added, of those tori defined by a rational relation of frequencies? It can be shown that resonances completely destroy resonant tori \cite{lichtenberg}. Consequently the trajectories on these tori --which, remember, in the absence of perturbation follow well defined closed and periodic curves on them-- lose their bounds and wander through the phase space (in fact, only through the subspace given by the  surface of constant energy). Therefore, motion on resonant tori becomes completely irregular and unstable. This is an example of the so-called {\it hard chaos}, as opposed to {\it soft-chaos} which relates to the motion near KAM surfaces \cite{gutzwiller} (see also next section).

\begin{figure}
\begin{center}
\includegraphics[scale=0.70]{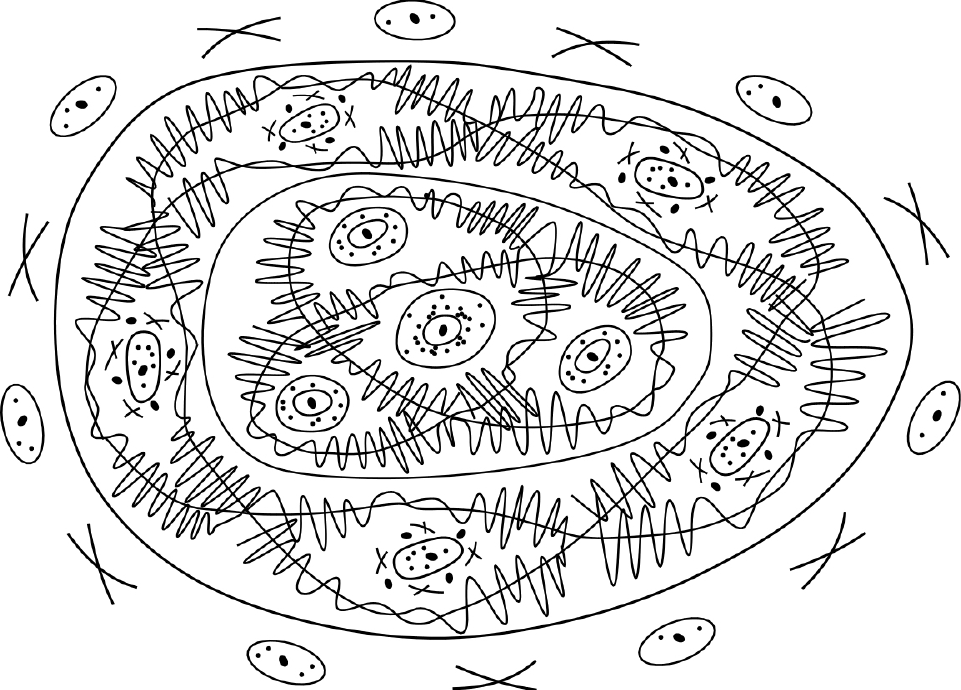} 
\caption{Schematic representation of Poincar\'e sections for nonintegrable systems stressing the complex mixture of regular and chaotic trajectories.}
\label{Fig9}
\end{center}
\end{figure}

Fortunately, as KAM theorem proves, the set of resonant tori, on which regular motion is destroyed, has a vanishingly small measure as long as the perturbation is small. Equivalently, the remaining slightly deformed tori where the motion is still regular, have a measure close to $1$ if $\epsilon$ is small. This means that most tori are not destroyed for small perturbations and KAM surfaces almost densely fill the energy shell and between them there appear ``islands'' with chaotic motion. This results in Poincar{\'e} sections which are 
extraordinarily complex and that some authors picture as in Fig. \ref{Fig9}.

\section{Miscellaneous results}

In this closing section we briefly outline some additional developments and applications of the theory.

\subsection{Arnold diffusion}

As KAM theorem proves, the energy shell of a conservative and nonintegrable Hamiltonian of the form $H=H_0+\epsilon H_1,$ where $H_0$ is integrable and $\epsilon$ is small, is densely filled with KAM surfaces on which phase-space trajectories are fairly regular. However, outside KAM tori trajectories are random. In order to analyze these chaotic trajectories we must distinguish among the cases of two and more than two degrees of freedom. 

Let us recall that, for integrable conservative systems with $n$ degrees of freedom, the phase space is $2n$ dimensional, the energy shell is $(2n-1)$ dimensional and invariant tori are $n$ dimensional. Hence, for $n=2$, the two dimensional tori are embedded in the three-dimensional energy surface. This means that any closed surface, such as a torus, divide the energy shell into an inner region and an outer region. The same applies to KAM tori of conservative nonintegrable system with $2$ degrees of freedom. Accordingly, any chaotic trajectory, located in the interspace between two tori, cannot escape from it. No matter how complicated its motion is, the trajectory never leaves the region and the action variables remain close to their initial values. This is another aspect of the soft-chaos mentioned above, for there exists certain stability of motion because KAM surfaces densely populate the energy shell and are, therefore, very close to each other.

\begin{figure}
\begin{center}
\includegraphics[scale=0.70]{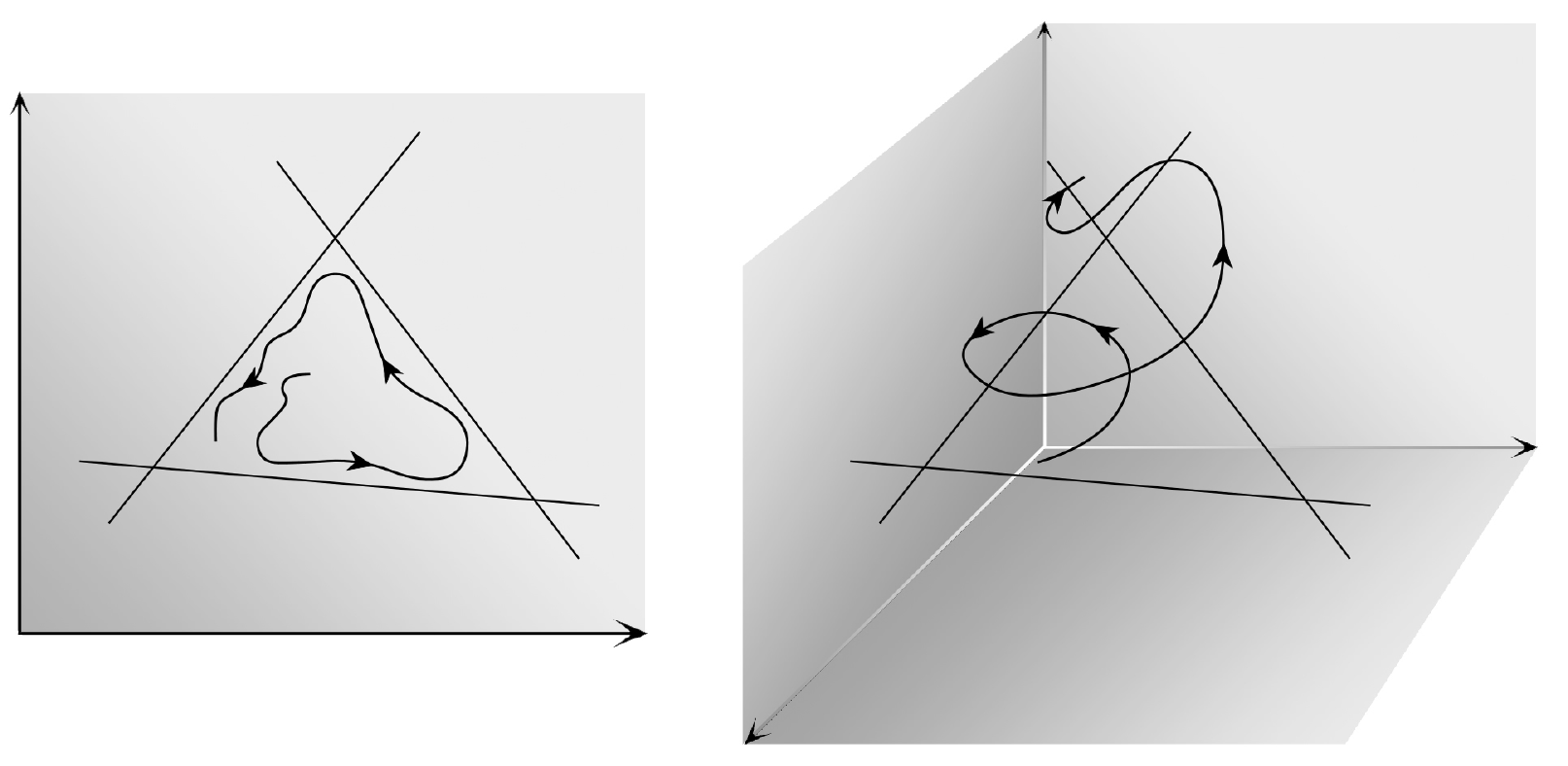} 
\caption{phase-space trajectories are confined by lines (KAM surfaces) in a two dimensional space but not in three dimensions.}
\label{Fig10}
\end{center}
\end{figure}

However, for $n\geq 3$ chaotic trajectories located in regions between tori can escape to other regions of the energy shell (see Fig. \ref{Fig10}). Thus, for instance, if $n=3$ the energy shell is $5$ dimensional, KAM surfaces $3$ dimensional, and chaotic trajectories have two additional dimensions to wander through. Hence, {\it for $n\geq 3$ the existence of KAM surfaces does not guarantee the stability of motion}, as two trajectories which are initially very close to each other can be very distant after a finite period of time. This phenomenon, that exists for arbitrarily small perturbations, is known as {\em Arnold diffusion} and only appears in systems with more than $2$ degrees of freedom.

The set of wandering trajectories resulting from it constitute the so-called {\it Arnold web}. There have been several attempts to determine the average speed of the spreading of trajectories defined as $\vert \Delta \mathbf{J}/\Delta t \vert$, where $\Delta \mathbf{J}=\mathbf{J}(t)-\mathbf{J}(0)$. Several theoretical models and numerical studies seem to indicate that this velocity is very small. Some models predict that \cite{lichtenberg}
$$
\vert \Delta \mathbf{J} \vert < \epsilon^{1/2} \qquad {\rm for} \qquad t<\epsilon^{-1}\exp\left(\epsilon^{-1/2} \right),
$$
which means, that for small $\epsilon$, the action at time $t$, $\mathbf{J}(t)$, remains close to its initial value, $\mathbf{J}(0)$, for a very long time  $t\sim\exp\left(\epsilon^{-1/2}\right)$. In other words, the average diffusion velocity is of the order
$$
\left|\frac{\Delta\mathbf{J}}{\Delta t}\right|
\sim e^{-1/\sqrt{\epsilon}},
$$
which is exponentially small and practically undetectable if $\epsilon\ll 1$.

\subsection{Planetary motion}

Let us return to the problem of the long-time stability of the solar system, one of the major challenges of mechanics for more than three centuries. As we know, the solar system is a nonintegrable many-body problem which, to first approximation, can be considered conservative. The key point is whether planetary motions are quasiperiodic or random. The solar system is too complicated and a direct application of the KAM theorem does not bring practical results. On the other hand, planet orbits exhibit great regularity and until recently it was thought that the solar system evolves quasiperiodically. In spite of this, there seems to be some evidence that this may not be the case. 

In 1857 the American astronomer D. Kirkwood discovered a series of discontinuities in the distribution of main-belt asteroids which circle the Sun and lie between Mars and Jupiter. There are regions where there is a depletion in the number of asteroids. These regions are called {\it Kirkwood gaps}. One of the most plausible explanation for them has been given by S. F. Dermott and C. D. Murray \cite{dermott}, who in 1983 proposed that the gaps are due to resonances in the three-body problem of the Sun, Jupiter, and the individual asteroids \cite{dermott,reichl}. That is, Jupiter perturbs the asteroid's almost Keplerian orbit around the Sun and creates resonances in the asteroid's phase space. The location of the observed gaps are found to be very close to the resonant surfaces of Jupiter and the asteroids, which would agree with KAM theory about the destruction of resonant tori. It has also been argued that the mechanism of Arnold diffusion might be responsible for removing asteroids from the gap regions and force some of them to collide with Mars \cite{lichtenberg}. 

Let us finally mention that there seems to be some numerical evidence that the long-term motion of the former planet Pluto is chaotic and also that of some of the main planets of the solar system \cite{reichl}. All these results seem to indicate that the long-time motion of the solar system may be chaotic and, to some extent, unpredictable.

\subsection{Quantum chaos}

Let us finish this exposition by giving a very brief view on the possible manifestation of chaotic behaviour in quantum systems. 

If one naively tries to extend, in a direct manner, the concepts of regular and chaotic trajectories to quantum systems, one is immediately faced with an unsurmountable difficulty: due to Heisenberg's uncertainty principle, between canonically conjugated variables, it makes no sense to define trajectories in the quantum phase space. 

In fact, the phase space of any system of $n$ degrees of freedom is divided in cells of volume $h^{n}$ ($h$ is Planck's constant) and one cannot determine at which point inside a given cell the system is. This renders the construction of phase-space trajectories meaningless. Therefore, the notion of invariant tori,  over which the trajectories of integrable systems stay, also loses its meaning in the quantum world. 

The problem of the quantization of nonintegrable Hamiltonians can be traced back to the early days of the old quantum theory \cite{gutzwiller} (see \cite{bergia} for a thorough review). However, strictly speaking, there is no such think as ``quantum chaos'', although there have been detected some peculiar effects in quantum systems whose classical counterparts correspond to chaotic systems.  Quoting M. Gutzwiller: ``the term `quantum chaos' serves more to describe a conundrum than to define a well-posed problem'' \cite{gutzwiller(b)}. 

What is understood by quantum chaos is essentially the relation between classical chaotic systems and their quantum counterparts. Let us recall that in quantum mechanics, the state of the system is not specified by a point $(\mathbf{q},\mathbf{p})$ of the phase space but by a wave function $\psi(\mathbf{q},t)$ giving the probability of finding the system at position $\mathbf{q}$ in time $t$. However, this wave-like character of quantum systems implies a smooth nature. How can the irregular character of classical chaos be reconciled with the smooth and wave-like nature of phenomena on the atomic scale governed by quantum mechanics? \cite{gutzwiller(b)}. What are, therefore, the quantum manifestations of chaos?.  

There are several places and situations where one can look for the ``quantum signatures'' of chaos \cite{haake}, and substantial research has been done along this direction in the last three decades. We can single out one aspect where quantum chaos is quite significant. The first place to look is  in the energy levels, not a particular level but the distribution  of them. Thus, and somewhat paradoxically, in nonchaotic quantum systems (i.e., those quantum systems whose classical counterparts are integrable) the energy levels are completely random without any  correlation, while in chaotic systems the energy levels show strong correlations between them, which implies a lower degree of randomness. There are other situations in which quantum manifestations of chaos appear as, for example, in the distribution of stationary states and in electron scattering, among others. We refer the interested reader to more specialized literature for further information (see, for instance, \cite{gutzwiller,haake,reichl}).

\section{Closing words}

There is a line of thought, among some philosophers and scientists, affirming that science is a construction, being scientists designers who fit as many data as possible within consistent frames. Opposite to that, others believe, and this is the traditional thinking, that science is discovery and the scientist is like a detective struggling to unveil the hidden truth \cite{esp}, ``the secret of the Old One'', as Einstein put it \cite{born-einstein}. Obviously, mainstream scientists lie somewhere in between these two extremes. 

Whatever the interpretation one may support, it is beyond any doubt that classical mechanics is one of the finest and most creative works that scientific endeavour can yield. In its most sophisticated form, the Hamiltonian approach, classical mechanics even arrives to forecast its own limitation: the long-time unpredictable behaviour of nonintegrable systems which, as mechanics also proves, are ubiquitous in nature. The solar system, galaxies, galaxy clusters, perhaps atoms and beyond, all of them are nonintegrable.

In this review we have tried to report on this striking consequence of classical mechanics. In a world of uncertainties --it all began with quantum mechanics \cite{bohm}-- we can add, as Poincar\'e envisaged, another one: the classical uncertainty.


\acknowledgments 

Partial financial support from the Ministerio de Ciencia e Innovaci\'{o}n under Contract No. FIS 2009-09689 is acknowledged. We thank Miquel Montero, Josep Perell\'o and Nitin Rughoonauth for a careful reading of the manuscript. We are indebted to Prof. Luis Navarro for useful discussions and suggestions.  

\appendix

\section{Integrability and the involution property}
\label{appenA0}

We know that, by definition, a system is integrable when there exist $n$ integrals of motion $I_i(\mathbf{q},\mathbf{p})=\alpha_i$ that are independent and in involution:
$$
[I_i,I_j]=0.
$$
This property allows us to use the $I$'s as the transformed momenta of some canonical transformation. Indeed, let us recall from (\ref{canonical}) that any set of $n$ variables $\mathbf{p'}=(p'_1,\dots,p'_n)$ can be considered as the momenta associated to some Hamiltonian if $[p'_i,p'_j]=0$. Accordingly, we make the canonical transformation
$$
(\mathbf{q},\mathbf{p})\longrightarrow (\mathbf{q'},\mathbf{p'}=\mbox{\boldmath$\alpha$}),
$$
in which the integrals of motion are the transformed momenta because the involution property grants it. The generating function of this canonical transformation is given by the indefinite integral
$$
S(\mathbf{q},\mbox{\boldmath$\alpha$})=\int\sum_{k=1}^n p_k(\mathbf{q},\mbox{\boldmath$\alpha$})dq_k,
$$
where $p_k(\mathbf{q},\mbox{\boldmath$\alpha$})$ are the momenta written in terms of the coordinates and the constants of motion. 

The equations of the canonical transformation are
$$
\mathbf{p}=\frac{\partial S}{\partial\mathbf{q}}, \qquad \mathbf{q'}=\frac{\partial S}{\partial\mbox{\boldmath$\alpha$}},
$$
and, since we deal with a time independent transformation, the transformed Hamiltonian, $H'=H(\mathbf{q'},\mbox{\boldmath$\alpha$})$, coincides with the original one written in terms of the new coordinates and momenta. 

In the new variables Hamilton equations read
$$
\mathbf{\dot{q}'}=\frac{\partial H}{\partial \mbox{\boldmath$\alpha$}}, \qquad 
\mbox{\boldmath$\dot{\alpha}$}=\frac{\partial H}{\partial \mathbf{q'}}.
$$
But $\mbox{\boldmath$\dot{\alpha}$}=0$, hence $\partial H/\partial{\mathbf{q'}}=0$ and $H$ is independent of the coordinates depending only upon the constants of motion:
$$
H=H(\alpha_1,\dots,\alpha_n)=E.
$$
Therefore, from the first group of Hamilton equations, we see that
$$
\mathbf{\dot{q}'}=\frac{\partial H(\mbox{\boldmath$\alpha$})}{\partial \mbox{\boldmath$\alpha$}}\equiv\mbox{\boldmath$\nu$}(\mbox{\boldmath$\alpha$})
=\ {\rm constant}
$$
and the equations of motion read
\begin{equation}
\mathbf{q'}(t)=\mbox{\boldmath$\nu$}(\mbox{\boldmath$\alpha$}) t+\ {\rm constant} \qquad {\rm and}\qquad 
\mathbf{p'}(t)=\mbox{\boldmath$\alpha$}=\ {\rm constant}.
\label{complete_solution}
\end{equation}
In this way we have completely integrated Hamilton equations for the motion of the system. Hence, the existence of $n$ independent integrals in involution leads to the complete integration of Hamilton equations. 

\section{Periodicity of the angle variables}
\label{appenA}

The variation, $\Delta_i\theta_j$, of the angle $\theta_j$ when the coordinate $q_i$ performs a complete oscillation is defined by
$$
\Delta_i\theta_j\equiv\sum_{k=1}^n\oint_{\Gamma_i}\frac{\partial\theta_j}{\partial q_k}dq_k,
$$
which, recalling that $\theta_j=\partial S/\partial J_j$ (see (\ref{tc})), we write as
$$
\Delta_i\theta_j=\sum_{k=1}^n\oint_{\Gamma_i}\frac{\partial^2 S}{\partial q_k\partial J_j}dq_k=\frac{\partial}{\partial J_j}\oint_{\Gamma_i}\sum_{k=1}^n\frac{\partial S}{\partial q_k}dq_k.
$$
But from (\ref{S}) we have $\partial S/\partial q_k=p_k$; hence
$$
\Delta_i\theta_j=\frac{\partial}{\partial J_j}\oint_{\Gamma_i}\sum_{k=1}^n p_k dq_k.
$$
Using the definition (\ref{2-2}) we get
$$
\Delta_i\theta_j=2\pi\frac{\partial J_i}{\partial J_j}
$$
and, since $(J_1,\dots,J_n)$ are independent variables, we finally obtain
$$
\Delta_i\theta_j=2\pi\delta_{ij},
$$
which is equation (\ref{delta_ij}).

\section{Commensurable frequencies}
\label{appenB}

From (\ref{q(t)}) we have
$$
q_j(t)=\sum_{\mathbf{k}}a^{(j)}_\mathbf{k} e^{i\mathbf{k}\cdot\mbox{\boldmath$\omega$}t},
$$
whence
$$
q_j(t+T_0)=\sum_{\mathbf{k}}a^{(j)}_\mathbf{k} e^{i\mathbf{k}\cdot\mbox{\boldmath$\omega$}t}e^{i\mathbf{k}\cdot\mbox{\boldmath$\omega$}T_0},
$$
and $q_j(t+T_0)=q_j(t)$ if and only if $e^{i\mathbf{k}\cdot\mbox{\boldmath$\omega$}T_0}=1$, which implies 
\begin{equation}
\mathbf{k}\cdot\mbox{\boldmath$\omega$}T_0=2\pi l,
\label{integer_1}
\end{equation}
where $l=0,\pm 1,\pm 2,\cdots$ is an arbitrary integer. This condition is equivalent to 
$$
\mbox{\boldmath$\omega$} T_0=2\pi\mathbf{l},
$$
where 
$$
\mathbf{l}=(l_1,\dots,l_n)
$$
$(l_j=0,\pm 1,\pm 2,\cdots)$ is an arbitrary integer vector. In effect, if, as (\ref{integer_1}) demands, the scalar product of {\it any} integer vector $\mathbf{k}$ with the real vector $(1/2\pi)\mbox{\boldmath$\omega$}T_0$ has to be an integer number, then necessarily $(1/2\pi)\mbox{\boldmath$\omega$} T_0$ has to be an integer vector. 

Therefore, any coordinate (and the same applies to momenta) will be a periodic function of time, that is to say, there will exist a single period $T_0$ for the entire system, if all frequencies $\mbox{\boldmath$\omega$}=(\omega_1,\dots,\omega_n)$ are integer multiples of a single frequency $\omega_0=2\pi/T_0$. That is
\begin{equation}
\mbox{\boldmath$\omega$}=\omega_0\mathbf{l},
\label{commen_1}
\end{equation}
which is equation (\ref{fundamental}) of the main text.

\section{Canonical perturbations}
\label{appenC}

In order to prove equation (\ref{H'}) we have to obtain the transformed Hamiltonian $H'$ under the canonical transformation (\ref{CT}). Since we are dealing with a time-independent transformation $H'=H$, and from (\ref{5-2}) we write
\begin{equation}
H'(\mbox{\boldmath$\theta'$}, \mathbf{J'})=H_0 \left(\mathbf{J}(\epsilon) \right)+ \epsilon\sum_{\mathbf{k}}H^{(1)}_{\mathbf{k}}(\mathbf{J}(\epsilon))\cos\bigl[\mathbf{k}\cdot\mbox{\boldmath$\theta$}(\epsilon)\bigr],
\label{C1}
\end{equation}
where $\mathbf{J}(\epsilon)$ and $\mbox{\boldmath$\theta$}(\epsilon)$ are given (implicitly) in terms of $\mathbf{J'}$ and $\mbox{\boldmath$\theta'$}$ by (\ref{J})--(\ref{theta}):
\begin{equation}
\mathbf{J}(\epsilon)=\mathbf{J'}+
\epsilon\sum_{\mathbf{k}} \mathbf{k} S_{\mathbf{k}}(\mathbf{J'})\cos\left[\mathbf{k}\cdot\mbox{\boldmath$\theta$}(\epsilon)\right]+
{\rm O}(\epsilon^2),
\label{J_A}
\end{equation}
\begin{equation}
\mbox{\boldmath$\theta'$}=\mbox{\boldmath$\theta$}(\epsilon)+\epsilon \sum_{\mathbf{k}} 
\frac{\partial S_{\mathbf{k}}(\mathbf{J'})}{\partial\mathbf{J'}}\sin\left[\mathbf{k}\cdot\mbox{\boldmath$\theta$}(\epsilon)\right]+
{\rm O}(\epsilon^2).
\label{theta_A}
\end{equation}

We will expand the right hand side of (\ref{C1}) in powers of $\epsilon$. Let us first observe, as trivially seen from (\ref{J_A}) and (\ref{theta_A}), that 
\begin{equation}
\mathbf{J}(\epsilon)=\mathbf{J'}+ O(\epsilon), \qquad \mbox{\boldmath$\theta$}(\epsilon)=\mbox{\boldmath$\theta'$}+ O(\epsilon),
\label{C2}
\end{equation}
hence
$$
H^{(1)}_{\mathbf{k}}(\mathbf{J}(\epsilon))=H^{(1)}_{\mathbf{k}}(\mathbf{J'})+O(\epsilon), \qquad  \cos\left[\mathbf{k}\cdot\mbox{\boldmath$\theta$}(\epsilon)\right]=\cos\left(\mathbf{k}\cdot\mbox{\boldmath$\theta'$}\right)+O(\epsilon).
$$
Therefore, (\ref{C1}) reads
\begin{equation}
H'(\mbox{\boldmath$\theta'$}, \mathbf{J'})=H_0 \left(\mathbf{J}(\epsilon) \right)+ \epsilon\sum_{\mathbf{k}}H^{(1)}_{\mathbf{k}}(\mathbf{J'})\cos\left(\mathbf{k}\cdot\mbox{\boldmath$\theta'$}\right)+O(\epsilon^2).
\label{C3}
\end{equation}

We next develop the unperturbed Hamiltonian $H_0 \left(\mathbf{J}(\epsilon) \right)$ in powers of $\epsilon$ up to first order; by Taylor expansion we have
\begin{equation}
H_0 \left(\mathbf{J}(\epsilon) \right)=H_0 \left(\mathbf{J}(0) \right)+\epsilon\left. \frac{d}{d\epsilon} H_0 \left(\mathbf{J}(\epsilon) \right)\right|_{\epsilon=0}+O(\epsilon^2),
\label{C4}
\end{equation}
where $\mathbf{J}(0)=\mathbf{J'}$ (see (\ref{C2})). On the other hand, using the chain rule, we write
$$
\frac{d}{d\epsilon} H_0 \left(\mathbf{J}(\epsilon) \right)=
\frac{\partial H_0}{\partial\mathbf{J}(\epsilon)}\cdot\frac{d\mathbf{J}(\epsilon)}{d\epsilon},
$$
where the dot means the scalar product; but from (\ref{J_A}) we get
$$
\frac{d\mathbf{J}(\epsilon)}{d\epsilon}=\sum_{\mathbf{k}} \mathbf{k} S_{\mathbf{k}}(\mathbf{J'})\cos\left[\mathbf{k}\cdot\mbox{\boldmath$\theta$}(\epsilon)\right]+O(\epsilon),
$$
hence,
$$
\left. \frac{d}{d\epsilon} H_0 \left(\mathbf{J}(\epsilon) \right)\right|_{\epsilon=0}=
\left. \frac{\partial H_0}{\partial\mathbf{J}(\epsilon)}\right|_{\epsilon=0}\cdot
\sum_{\mathbf{k}} \mathbf{k} S_{\mathbf{k}}(\mathbf{J'})\cos\left(\mathbf{k}\cdot\mbox{\boldmath$\theta'$}\right),
$$
since $\mbox{\boldmath$\theta$}(0)=\mbox{\boldmath$\theta'$}$. 

Recalling that $\partial H_0/\partial\mathbf{J}=\mbox{\boldmath$\omega$}$ are the frequencies of the unperturbed Hamiltonian, we have
$$
\left.\frac{\partial H_0}{\partial\mathbf{J}(\epsilon)}\right|_{\epsilon=0}=
\mbox{\boldmath$\omega$}(\mathbf{J}(0))=\mbox{\boldmath$\omega$}(\mathbf{J'}).
$$
Therefore,
$$
\left. \frac{d}{d\epsilon} H_0 \left(\mathbf{J}(\epsilon) \right)\right|_{\epsilon=0}=
\sum_{\mathbf{k}} \mathbf{k}\cdot\mbox{\boldmath$\omega$}(\mathbf{J'}) S_{\mathbf{k}}(\mathbf{J'})\cos\left(\mathbf{k}\cdot\mbox{\boldmath$\theta'$}\right),
$$
which substituted into expansion (\ref{C4}) yields
$$
H_0 \left(\mathbf{J}(\epsilon) \right)=H_0 \left(\mathbf{J'}\right)+
\epsilon \sum_{\mathbf{k}} \mathbf{k}\cdot\mbox{\boldmath$\omega$}(\mathbf{J'}) S_{\mathbf{k}}(\mathbf{J'})\cos\left(\mathbf{k}\cdot\mbox{\boldmath$\theta'$}\right)+O(\epsilon^2);
$$
plugging this expression into (\ref{C3}) we get
$$
H'(\mbox{\boldmath$\theta'$}, \mathbf{J'})=H_0(\mathbf{J'})+\epsilon\sum_{\mathbf{k}}\left[H^{(1)}_{\mathbf{k}}(\mathbf{J'})+
\mathbf{k}\cdot\mbox{\boldmath$\omega$}(\mathbf{J'})
S_{\mathbf{k}}(\mathbf{J'})\right]\cos\left(\mathbf{k}\cdot\mbox{\boldmath$\theta$}\right)+{\rm O}(\epsilon^2),
$$
and, removing from the sum the term with $k_1=\cdots=k_n=0$, we finally have
$$
H'(\mbox{\boldmath$\theta'$}, \mathbf{J'})=H_0(\mathbf{J'})+\epsilon H_{\mathbf{0}}^{(1)}(\mathbf{J'})+
\epsilon\sum_{\mathbf{k}\neq\mathbf{0}}\left[H^{(1)}_{\mathbf{k}}(\mathbf{J'})+\mathbf{k}\cdot\mbox{\boldmath$\omega$}(\mathbf{J'})
S_{\mathbf{k}}(\mathbf{J'})\right]\cos\left(\mathbf{k}\cdot\mbox{\boldmath$\theta$}\right)+{\rm O}(\epsilon^2),
$$
which is equation (\ref{H'}).


\end{document}